\documentclass[superscriptaddress,aps,prl,reprint,twocolumn,amsmath,amssymb]{revtex4-2}
\usepackage{graphicx}
\usepackage{hyperref}
\usepackage{amssymb}
\usepackage{slashed}
\usepackage{dcolumn}
\usepackage{amsmath}
\usepackage{bm}
\usepackage{colordvi}
\usepackage{algorithm}
\usepackage{algpseudocode}
\usepackage{multirow}
\usepackage{titlesec}
\usepackage{newtxtext,newtxmath}
\usepackage{dsfont}
\usepackage{xcolor}
\usepackage{mathbbol}

\usepackage{hyperref}
\hypersetup{
    colorlinks=true,
    linkcolor=blue,
    citecolor=blue,     
    urlcolor=blue,
}

\allowdisplaybreaks

\usepackage{mathrsfs}
\makeatletter

\newcommand{\Rmnum}[1]{\expandafter\@slowromancap\romannumeral #1@}
\makeatother

\begin{document}

\title{Hidden Crossover and Relaxor-Like Response from Emerging Polar Skyrmion Correlations in Ferroelectric Superlattices}

\author{Zhi-Yang Wang}

\affiliation{Department of Materials Science and Engineering and Materials Research Institute, The Pennsylvania State University, University Park, PA 16802, USA}

\author{Fei Yang}
\email{fzy5099@psu.edu}

\affiliation{Department of Materials Science and Engineering and Materials Research Institute, The Pennsylvania State University, University Park, PA 16802, USA}

\author{Long-Qing Chen}
\email{lqc3@psu.edu}

\affiliation{Department of Materials Science and Engineering and Materials Research Institute, The Pennsylvania State University, University Park, PA 16802, USA}

\date{\today}

\begin{abstract}
Polar skyrmions in ferroelectric superlattices are nanoscale topological polarization textures typically regarded as weakly coupled objects confined to individual layers, with a role secondary to that of the underlying symmetry-breaking order parameter. Here using large-scale phase-field simulations of ferroelectric superlattices, we uncover a hidden thermal crossover deep inside the ferroelectric phase, where polar skyrmions evolve from an uncorrelated, layer-resolved state into an interlayer-correlated ensemble. This crossover occurs without additional symmetry breaking or a new order parameter, but produces a pronounced broad peak in the dielectric susceptibility. The anomaly originates from the competition between correlation-enhanced response, associated with the growth of interlayer skyrmion correlations, and polarization-induced stiffness, which suppresses dielectric fluctuations at low temperature. Under AC driving, the peak shifts with frequency, resembling relaxor ferroelectrics despite the absence of quenched disorder or polar nanoregions. Our results establish a disorder-free route to relaxor-like dielectric response and identify topological defect correlations as an organizing principle for thermodynamic anomalies, providing a mechanism distinct from conventional critical behavior associated with symmetry breaking and divergent order-parameter fluctuations.
\end{abstract}

\maketitle  

{\sl Introduction.---}Macroscopic thermodynamic anomalies in ordered phases are commonly understood through the Landau paradigm, in which thermodynamic responses are governed by symmetry breaking, the associated order parameter~\cite{landau2013course}, and its various smooth fluctuations~\cite{nambu1960quasi,goldstone1961field,goldstone1962broken,nambu2009nobel,yang2019gauge,ambegaokar1961electromagnetic,littlewood1981gauge,PhysRevB.104.214510,gpbp-qhp9,pekker2015amplitude}. This framework has been remarkably successful in describing ferroelectric transitions, where the dielectric susceptibility is tied to the onset and growth of spontaneous polarization. 
However, ordered phases can also host topological defects with well-defined topological charges, i.e., singular configurations of the order parameter that cannot be removed by smooth deformations~\cite{mermin_topological_1979}, such as vortices in superconductors~\cite{abrikosov_magnetic_1957,hess_scanning-tunneling-microscope_1989,blatter_vortices_1994-1}, skyrmions in magnetic systems~\cite{rosler_spontaneous_2006-3,muhlbauer_skyrmion_2009-2,Nagaosa2013}, and disclinations in liquid crystals~\cite{stephen_physics_1974,alexander_colloquium_2012}. These defects act as emergent objects with their own interactions, correlations, and dynamics, introducing a layer of collective behavior beyond that captured by the conventional order parameter. 
Their collective organization can dominate macroscopic responses: proliferation, binding, or unbinding of defects can control phase coherence and transport properties even in the absence of a conventional symmetry-breaking transition, as exemplified by vortex physics in superconductors~\cite{abrikosov_magnetic_1957,blatter_vortices_1994-1} and the Berezinskii-Kosterlitz-Thouless transition~\cite{kosterlit_ordering_nodate,kosterlitz_critical_1974-1,kosterlitz2016kosterlitz}. 
These examples illustrate that collective phenomena in many-body systems can be governed not only by symmetry breaking, but also by the topology and correlations of emergent defects that may dominate thermodynamic response deep inside an ordered phase~\cite{3q5l-46h5,xy6z-hxcv,b6vp-zt8z}.

Polar skyrmions are topological defects in ferroelectric systems~\cite{Paillard2025,Das2021,Wang2024,Gong2023,Li2025}, characterized by a quantized skyrmion number~\cite{junquera_topological_2023-2,Nagaosa2013},
\begin{equation}
    N_{sk}=\frac{1}{4\pi}\int d^2 \vec{r} \; \vec{p}\cdot\left(\frac{\partial\vec{p}}{\partial{x}}\times\frac{\partial\vec{p}}{\partial{y}}\right),
\end{equation}
here $\vec{p}$ denotes the normalized polarization vector field. Recent advances in atomic-resolution imaging have established their existence in ferroelectric superlattices, such as PbTiO$_3$/SrTiO$_3$ (PTO/STO)~\cite{das_observation_2019}. These topological textures exhibit unconventional functionalities, including local negative dielectric response~\cite{Das2021}, structural chirality~\cite{mccarter_structural_2022}, and terahertz collective dynamics~\cite{Li2025}, highlighting their potential as a platform for emergent mesoscale physics. A central open question is whether polar skyrmions can develop collective correlations beyond their conventional description as weakly interacting, layer-confined topological textures~\cite{ren_emergence_2024,geng_dipolar_2025}, while they are often treated as isolated topological excitations. This raises a more fundamental issue: how are such correlations encoded in the thermodynamic response of a material? In particular, it remains unclear whether the evolution from uncorrelated to correlated skyrmion states corresponds to a genuine symmetry-breaking phase transition, or instead reflects a correlation-driven crossover with distinct signatures in measurable quantities such as the dielectric susceptibility.

Driving a temperature-induced crossover between uncorrelated and 
coherently coupled polar skyrmions requires a superlattice in which the 
dielectric spacer develops a strongly temperature-dependent polarization. 
In such structures, dipolar fields originating from the ferroelectric 
layers can penetrate the spacer once it becomes sufficiently polarized, 
thereby enabling interlayer skyrmion coupling~\cite{wang_reversible_2023,wang_tuning_2024}. Here we show that polar skyrmions in oxide superlattices can develop strong interlayer correlations upon cooling, giving rise to a hidden thermal crossover that does not involve symmetry breaking or the emergence of a new order parameter, but instead reflects a collective reorganization of defect correlations and  strongly modifies the  dielectric response. 
Using large-scale phase-field simulations of prototypical 
(Pb$_x$Sr$_{1-x}$TiO$_3$)/(PbTiO$_3$) superlattices, we find that skyrmions evolve upon cooling from an uncorrelated layer-resolved state into an interlayer-correlated state. 
This crossover originates from the gradual buildup of 
interlayer correlations: while skyrmions remain effectively decoupled at 
high temperature, cooling enhances the polarization of the spacer layer, 
which activates dipolar-mediated coupling and drives the formation of 
correlated skyrmion stacks. 
The crossover produces a broad dielectric susceptibility peak well below the ferroelectric transition temperature. 
Unlike a critical anomaly, this peak results from the competition between the growth of skyrmion correlations and the increasing stiffness of the polarized state.  Remarkably, under AC driving the peak shifts with frequency, giving rise to a relaxor-like response in a structurally ordered system. 
These results establish topological defect correlations as a route to thermodynamic anomalies and relaxor-like dynamics beyond conventional symmetry-breaking and disorder-based mechanisms.

{\sl Model---}To investigate the thermodynamic properties of the (Pb$_x$Sr$_{1-x}$TiO$_3$)$_{16}$/(PTO)$_{16}$ superlattices across a range of temperatures, we employ the phase-field method. The polarization vector $\vec{P}(\vec{r}) = (P_1, P_2, P_3)$ is taken as the order parameter to construct the total free energy of the system~\cite{ref2}:
\begin{equation}
    F = \int_V \left[ f_{\text{L}}(\vec{P}) + f_{\text{g}}(\nabla \vec{P}) + f_{\text{elas}}(\vec{P}, \varepsilon_{ij}) + f_{\text{elec}}(\vec{P}, \vec{E}) \right] dV,
\end{equation}
where the four terms correspond to Landau-Devonshire bulk energy, gradient energy, elastic energy, and electrostatic
energy, respectively. The Landau-Devonshire energy density is expanded to the sixth order in $\vec{P}$ to capture the ferroelectric phase transitions:
\begin{equation}
    f_{\text{L}} = \alpha_i P_i^2 + \alpha_{ij} P_i^2 P_j^2 + \alpha_{ijk} P_i^2 P_j^2 P_k^2,
\end{equation}
where the quadratic coefficient $\alpha_i(T)$ follows a Curie-Weiss temperature dependence, while higher-order coefficients $\alpha_{ij}$ and $\alpha_{ijk}$ are taken to be temperature-independent.

Spatial variations are penalized by the gradient term,
\begin{equation}
    f_{\text{g}} = \frac{1}{2} G_{ijkl} \frac{\partial P_i}{\partial x_j} \frac{\partial P_k}{\partial x_l}, 
\end{equation}
which is assumed here to take the isotropic form: $g (\partial_j P_i)^2$, representing the nearest-neighbor interactions. Elastic effects are incorporated through electrostrictive coupling, 
\begin{equation}
    f_{\text{elas}} = \frac{1}{2} C_{ijkl} e_{ij} e_{kl} = \frac{1}{2} c_{ijkl} (\varepsilon_{ij} - \varepsilon^0_{ij}) (\varepsilon_{kl} - \varepsilon^0_{kl}),
\end{equation}
 with the spontaneous strain $\varepsilon^0_{ij} = Q_{ijkl} P_k P_l$. The mechanical equilibrium is enforced via $\partial_j \sigma_{ij}=0$ under thin-film boundary conditions~\cite{ref6}. Here, $C_{ijkl}$ is the elastic stiffness, 
 $e_{ij}$ is the elastic strain, $Q_{ijkl}$ represents the electrostrictive coefficients,  and $\sigma_{ij}$ is the elastic stress.

The electrostatic contribution is given by
\begin{equation}
    f_{\text{elec}} = -\frac{1}{2}\epsilon_0 \kappa_b E_i^2 - E_iP_i,
\end{equation}
where the electric field $\vec{E}$ is obtained by solving the Poisson equation $\nabla \cdot (\epsilon_b \vec{E} + \vec{P}) = 0$ assuming zero free charge density and short circuit boundary condition\cite{ref7}, with $\kappa_b$ being the isotropic background dielectric constant and $\epsilon_0$ being the vacuum dielectric permittivity.

\begin{figure}[b]
  {\includegraphics[width=8.7cm]{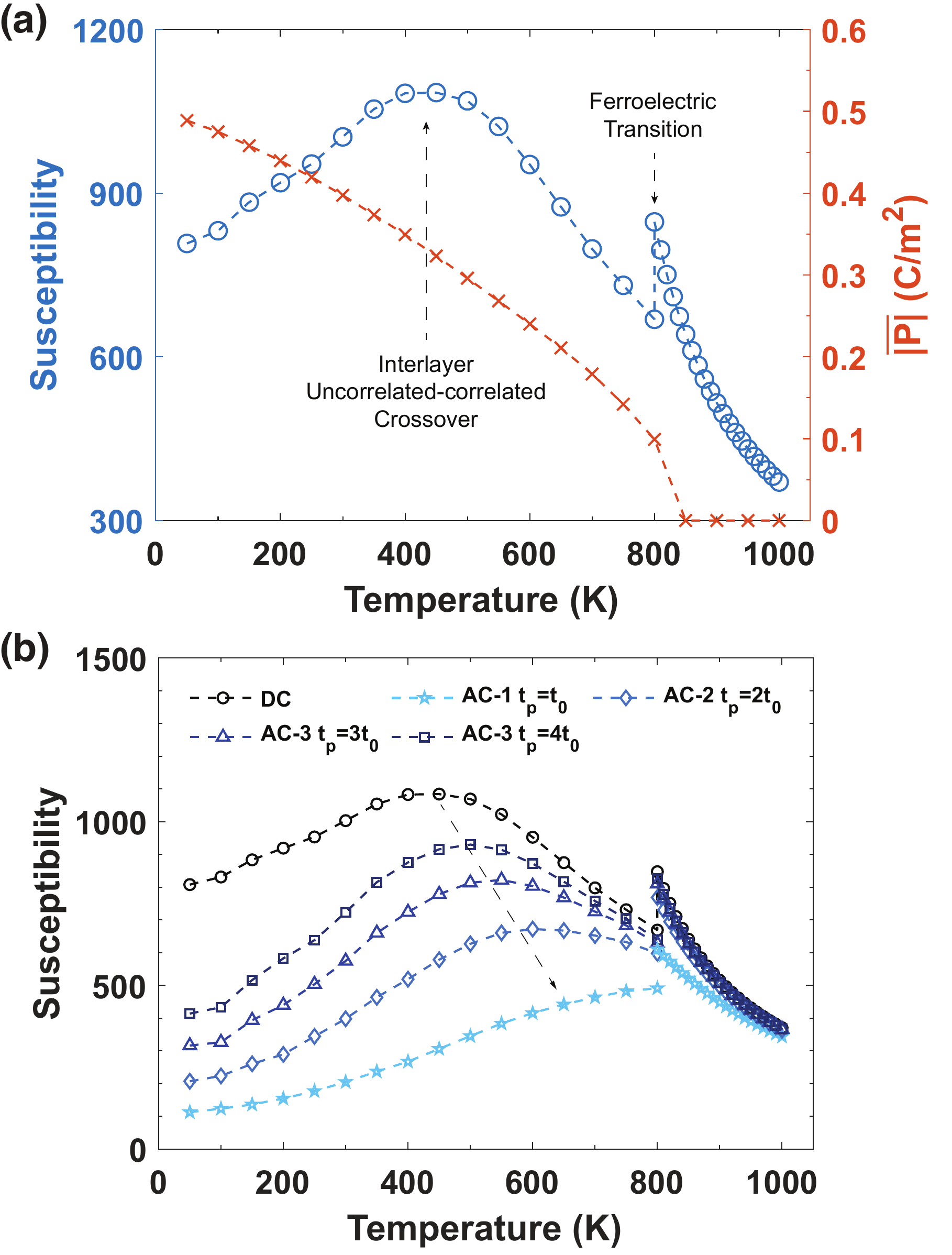}}
\caption{
(a) DC dielectric susceptibility and {spatially averaged polarization magnitude $\overline{P}$} as functions of temperature for the (Pb$_x$Sr$_{1-x}$TiO$_3$)$_{16}$/(PbTiO$_3$)$_{16}$ superlattice at $x=0.3$. (b) AC and DC dielectric susceptibilities versus temperature of the same superlattice. The period $t_p$ of the applied AC electric field is an integer multiple of the fundamental period $t_0$ (see  Supplementary Material). The black arrow indicates the direction of the dielectric-peak shift.
}
\label{fig1}
\end{figure}

The kinetics of the polarization is governed by the time-dependent Ginzburg-Landau (TDGL) equation:
\begin{equation}
    \frac{\partial P_i(\vec{r}, t)}{\partial t} = -\Gamma \frac{\delta F}{\delta P_i(\vec{r}, t)},
\end{equation}
which relaxes the system toward thermodynamic equilibrium. Here, $\Gamma$ is a kinetic coefficient associated with domain-wall mobility. Simulations are performed on a $200\Delta \times 200\Delta \times 200\Delta$ grid with $\Delta=0.4~\text{nm}$. The superlattice is constructed along the $z$ direction, consisting of a substrate (20 grids), a central superlattice region (176 grids), and a top {vacuum} layer (4 grids). The superlattice comprises alternating (PTO)$_{16}$ and (Pb$_x$Sr$_{1-x}$TiO$_3$)$_{16}$ layers, repeated five times, starting from a (Pb$_x$Sr$_{1-x}$TiO$_3$)$_{16}$ block. The parameters of the free-energy density function~\cite{ref3}, together with full numerical details and boundary conditions, are provided in the Supplemental Material~\cite{supple}, including Ref.~\cite{ref1,ref2,ref3,shirokov2022thermodynamic,ref6,ref7,ref5,chen1998applications,ref11,ref12,ref8,ref9,ref10,yang2020theory,shimano2020higgs,matsunaga2013higgs,matsunaga2014light,yang2024optical,yang2023optical,PhysRevB.106.L081105,Abid2021}.

{\sl Results---}We first focus on a representative composition, $x=0.3$, and compute the dielectric susceptibility as a function of temperature. As shown in Fig.~\ref{fig1}, a sharp discontinuous peak appears at $T_c\sim800$ K, signaling a first-order ferroelectric transition. At this transition, the cubic symmetry is broken in both the PTO and Pb$_x$Sr$_{1-x}$TiO$_3$ layers through a proximity effect, and {a finite local ferroelectric polarization, or equivalently, the spatially averaged polarization magnitude, $
\overline{P}=\frac{1}{V}\int_V |\mathbf{P}(\mathbf{r})|d\mathbf{r}$, emerges throughout the superlattice despite the nearly vanishing macroscopic polarization
 $\overline{\mathbf P}=\frac{1}{V}\int_V \mathbf P(\mathbf r)d\mathbf r
\approx 0$.} The associated dielectric anomaly at $T_c$ follows the conventional critical behavior of a ferroelectric transition, governed by symmetry breaking and enhanced order-parameter fluctuations. {The superlattice exhibits nearly complete compensation between upward and downward polarization domains, consistent with the previous studies~\cite{Nahas2020,Lupi2023,PhysRevLett.133.066801}.}

In conventional ferroelectrics (e.g., PbTiO$_3$~\cite{yang2024first,haun1987thermodynamic,remeika1970growth,ikegami1971electromechanical}), 
the dielectric susceptibility decreases monotonically upon cooling below 
the transition temperature. In contrast, in the (Pb$_x$Sr$_{1-x}$TiO$_3$)/(PbTiO$_3$) superlattice, a second broad peak emerges at 
$\sim 500$~K within the ferroelectric phase. The polarization varies 
smoothly across this temperature without any discontinuity, excluding an 
additional symmetry-breaking transition and identifying this feature as 
a crossover. The presence of this broad peak therefore signals a 
qualitative reorganization within the ferroelectric phase, which we 
attribute to the onset of correlated polar skyrmion behavior.

Remarkably, under an applied AC electric field, the susceptibility peak shifts toward higher temperature with increasing driving frequency, reminiscent of relaxor-like dielectric behavior~\cite{Bokov2006}. 
This resemblance is striking because the present superlattice is structurally ordered and contains no quenched disorder or polar nanoregions. 
Therefore, the broad dispersive response cannot be attributed to the conventional relaxor mechanism~\cite{Bokov2006}. 
Instead, it indicates the slow collective relaxation of an emerging interlayer-correlated skyrmion ensemble. 
The frequency-dependent shift thus provides dynamical evidence that the dielectric anomaly is controlled by defect correlations rather than by a conventional ferroelectric transition.

\begin{widetext}
  \begin{center}
\begin{figure}[htb]
  {\includegraphics[width=\textwidth]{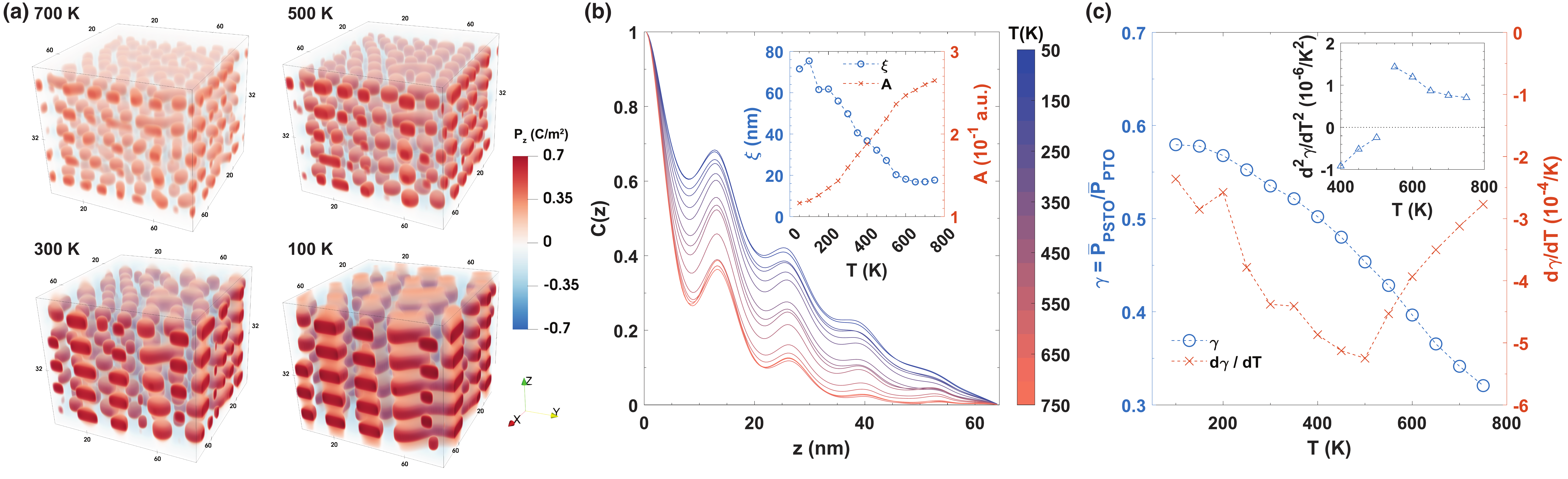}}
\caption{\textbf{(a)} Polar structures of the (Pb$_x$Sr$_{1-x}$TiO$_3$)$_{16}$/(PTO)$_{16}$   superlattice (x=0.3) at T = 700~K, 500~K, 300~K, and 100~K, illustrating the progressive spatial organization of polar textures upon cooling.  \textbf{(b)} Temperature dependence of the out-of-plane skyrmion correlation function, the correlation length (blue circles in the inset), and the oscillation amplitude (orange crosses in the inset). The colors of the correlation-function curves indicate temperature. \textbf{(c)} Polarization ratio $\gamma$ between Pb$_x$Sr$_{1-x}$TiO$_3$ and PTO layers (blue circles), together with its first derivative (orange crosses) and second derivative (inset) with respect to $T$. }
\label{fig2}
\end{figure}
\end{center}
\end{widetext}

To elucidate the origin of this crossover, we examine the evolution of the polar structure at representative temperatures (700~K, 500~K, 300~K, and 100~K), as shown in Fig.~\ref{fig2}(a). Below the ferroelectric transition ($\sim 800$ K), a finite polarization develops and polar skyrmions emerge in the PTO layers. At high temperatures, these skyrmions are essentially uncorrelated and randomly distributed across different layers. Upon cooling, however, they progressively align and form correlated structures along the out-of-plane direction, indicating the emergence of interlayer correlation. To quantify this behavior, we compute the out-of-plane autocorrelation function of the normalized polarization field $\vec{p}=\vec{P}/|\vec{P}|$,
\begin{equation}
    C(z)=\frac{1}{V} \int_V\;\vec{p}(x',y',z')\cdot\vec{p}(x',y',z'+z)dx'dy'dz'.
\end{equation}

The correlation function in Fig.~\ref{fig2}(b) exhibits a damped oscillatory behavior with an additional linear decay. The oscillatory component reflects the periodic arrangement of polar skyrmions across different PTO layers, while the exponential envelope captures the finite interlayer correlation length. The linear decay originates from a finite-size geometric effect, arising from the reduced overlap volume at large separations.

As the temperature decreases, the decay of the correlation function becomes progressively slower, indicating an increasing out-of-plane correlation length. To quantify this evolution, we fit the data using an exponentially damped harmonic form with a finite-size correction,
\begin{equation}
C(z)=A\Big[\cos\Big(\frac{2\pi z}{\lambda_0}+\phi\Big)+C_0\Big]\exp\Big(-\frac{z}{\xi}\Big)\left(1-\frac{z}{L}\right),
\end{equation}
where $A$ is the oscillation amplitude, $\lambda_0=12.8~\mathrm{nm}$ corresponds to the superlattice periodicity, $\xi$ denotes the out-of-plane correlation length of the superlattice{, and $C_0$ is a constant offset to account for finite-size and background effects.}  Such a form is characteristic of a correlated but non-long-range-ordered state with finite coherence length.  The geometry factor $(1-z/L)$ accounts for the finite thickness of the system.  The extracted temperature dependence of $\xi$ and $A$ is shown in the inset of Fig.~\ref{fig2}(b). Upon cooling, $\xi$ increases monotonically, demonstrating the progressive buildup of interlayer skyrmion correlations. In contrast, the oscillation amplitude $A$ decreases, indicating that the layer-resolved modulation becomes weaker as skyrmion textures in different PTO layers evolve toward a more correlated interlayer configuration.

The emergence of correlations is often associated with the development of collective degrees of freedom. The interlayer correlations of polar skyrmions here are controlled by the polarization of Pb$_x$Sr$_{1-x}$TiO$_3$ layers, which mediate the coupling between the adjacent PTO layers {(see Fig.~SII)}. At high temperatures, the weak polarization of the Pb$_x$Sr$_{1-x}$TiO$_3$ layers effectively decouples the neighboring PTO layers, resulting in uncorrelated, layer-resolved skyrmion distributions. Upon cooling, the polarization of the Pb$_x$Sr$_{1-x}$TiO$_3$ layers increases rapidly, enabling dipolar interactions to propagate across the spacer more effectively and thereby enhancing interlayer coupling. The enhanced coupling promotes a coherent alignment of skyrmions across different layers, which reduces the polarization gradient energy and stabilizes the correlated state.

To quantify this mechanism, we introduce the polarization ratio $\gamma = \bar{P}_{\text{Pb$_x$Sr$_{1-x}$TiO$_3$}}/\bar{P}_{\mathrm{PTO}}$ [Fig.~\ref{fig2}(c)], together with its first and second temperature derivatives. As temperature decreases, $\gamma$ grows monotonically with a crossover from rapid increase to saturation, reflecting the strengthening of interlayer coupling.  Notably, near $T \sim 500~\mathrm{K}$, the curvature of $\gamma(T)$ changes sign, i.e., $d^2\gamma/dT^2=0$, signaling an inflection point. This temperature coincides with the broad peak in the dielectric susceptibility below the ferroelectric transition, providing a direct link between the evolution of interlayer coupling and the thermodynamic response. We therefore identify this inflection point as the characteristic crossover temperature.

The correlation length $\xi$ reflects the strength of the effective interlayer coupling, which is controlled by the polarization ratio $\gamma$, and increases upon cooling. Theoretically, the dielectric response is enhanced by the growth of the correlated volume associated with skyrmion stacking, but suppressed by the increasing stiffness of the polarized state. 
{As demonstrated by a developed effective analytical model in Sec.~SVI~A}, the susceptibility is controlled by two competing tendencies: an increasing correlation scale, represented by $\xi(T)$, and an effective polarization stiffness, denoted by $K(T)$. 
The broad maximum in $\chi(T)$ emerges when the enhancement from growing skyrmion correlations is overtaken by the stiffening of the ferroelectric background at lower temperature. 
This competition gives rise to a broad susceptibility peak well below the ferroelectric transition, revealing a hidden crossover governed by defect correlations rather than the symmetry breaking described by the conventional Landau paradigm of phase transitions.

\begin{figure}
  {\includegraphics[width=8.2cm]{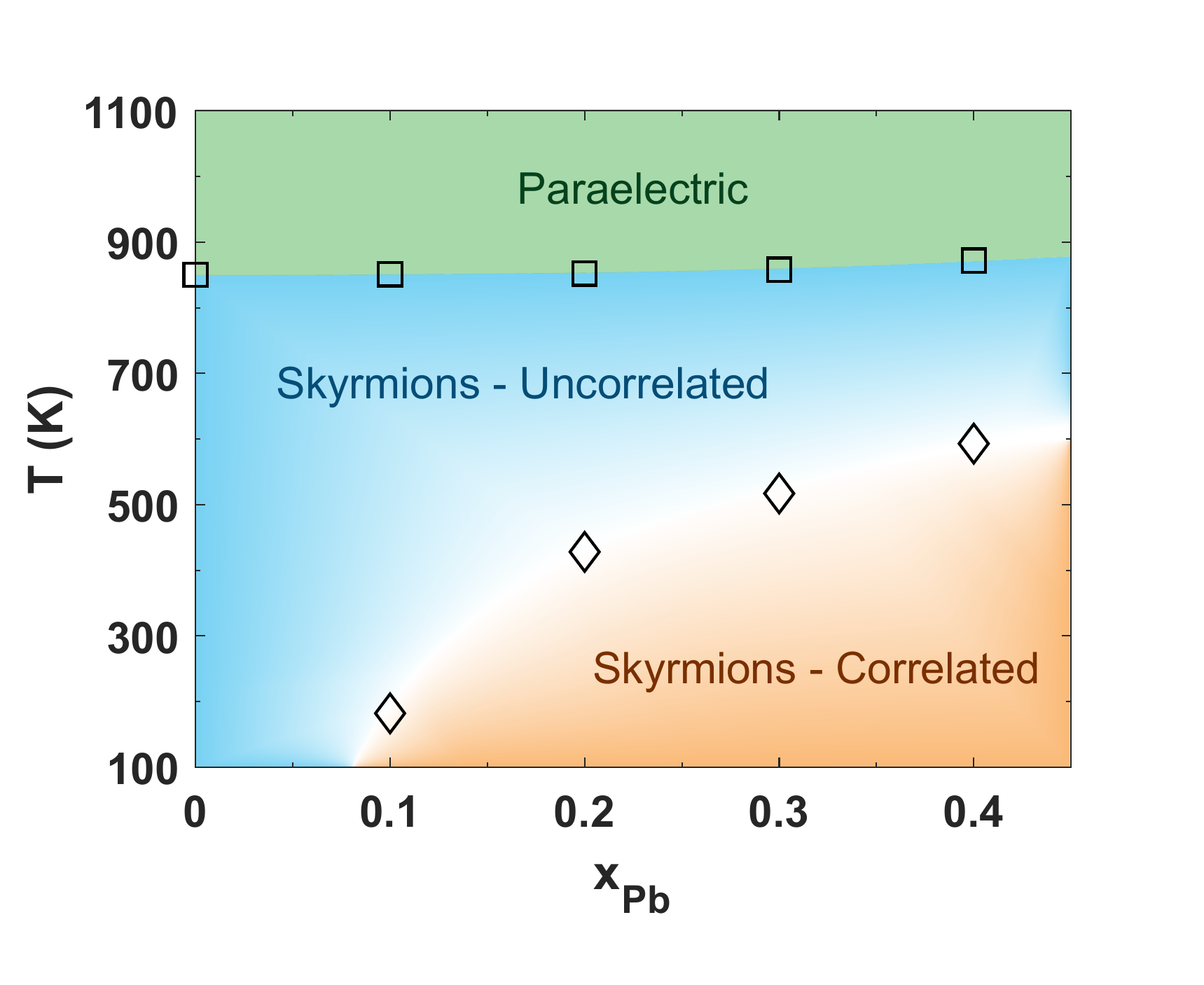}}
\caption{The calculated temperature-composition regime map of (Pb$_x$Sr$_{1-x}$TiO$_3$)$_{16}$/(PTO)$_{16}$ superlattices, {obtained during
continuous cooling. The phase boundaries correspond specifically
to the cooling path.} 
The green, blue, and orange regions denote the paraelectric regime, uncorrelated skyrmion regime, and interlayer-correlated skyrmion regime, respectively. 
Diamonds mark the crossover temperature between the uncorrelated and correlated skyrmion regimes, while squares denote the ferroelectric transition temperature extracted from simulations. 
The color shading is a guide to the eye.
}
\label{fig3}
\end{figure}

Increasing the Pb concentration in the Pb$_x$Sr$_{1-x}$TiO$_3$ layers enhances their polarization, thereby strengthening the interlayer coupling between skyrmions in adjacent PTO layers. This leads to a pronounced compositional dependence of the crossover temperature between uncorrelated and correlated skyrmion states. In Fig.~\ref{fig3}, we construct a temperature-composition regime map for the present superlattice. For low Pb concentration ($x \lesssim 0.1$), skyrmions remain effectively uncorrelated over the entire temperature range. At higher concentrations, a crossover emerges, with the crossover temperature (diamonds) increasing monotonically with $x$, reflecting the enhanced interlayer coupling. The ferroelectric transition temperature $T_c$ (squares) exhibits a weak dependence on composition, since the transition is  dominated by the PTO layers where $T_c$ is substantially higher than that of Pb$_x$Sr$_{1-x}$TiO$_3$ in the composition range ($0 \le x \le 0.4$).

{\sl Discussion.---}Our results {suggest that the emerging interlayer correlation of skymions is an intrinsic low-temperature property of the system~(See Sec.~SV)}, and demonstrate that thermodynamic responses in ferroelectric superlattices can be governed by correlations among topological defects rather than by symmetry breaking alone. The crossover identified here does not introduce a new order parameter, but reflects a collective reorganization of skyrmions from an uncorrelated layer-resolved state into an interlayer-correlated ensemble. This establishes a distinct mechanism where thermodynamic anomalies arise from the evolution of topological-defect correlations, rather than from a true thermodynamic phase transition.

A central consequence is the emergence of a pronounced and broad dielectric susceptibility peak deep within the ordered phase. This feature originates from the competition between correlation-enhanced response and polarization-induced stiffness.  The susceptibility maximum appears when these two tendencies balance, providing a mechanism distinct from conventional critical behavior tied to symmetry breaking and divergent order-parameter fluctuations. Another important implication is the emergence of relaxor-like dynamics without quenched disorder. Conventional relaxor ferroelectrics are commonly associated with chemical disorder and polar nanoregions, whereas the present superlattice is structurally ordered. 
The frequency-dependent shift of the dielectric peak here originates from the collective dynamics of correlated polar skyrmions {(See~Sec.~SVI~B for details)}, establishing a disorder-free route to broad, dispersive dielectric response.

{We note that the relaxor-like behavior previously reported in 
$(\mathrm{BaTiO}_3)_n/(\mathrm{SrTiO}_3)_n$ superlattices~\cite{Lupi2023}
shares several phenomenological similarities with our results, including frequency dispersion of the dielectric peaks. Its relaxor-like response was attributed to
thermally induced width modulation of domains and easily accessible fluctuations of multiple dipolar
textures, including antipolar stripe domains, diagonal domain configurations,
and slush-like polar states. Thus, even without quenched chemical disorder,
 multiple metastable configurations provide a glassy-like
configurational landscape that supports collective slowing down and
Vogel-Fulcher-type dynamics. In the present system, however, the topological
character of the skyrmions strongly constrains configurational fluctuations
among distinct metastable textures, since, unlike ordinary dipolar textures
that can continuously reorganize and transform through fluctuations, the skyrmions are constrained
to preserve their topological character,   thereby shifting the dominant collective
dynamics from fluctuations among competing dipolar textures to the formation
and dissolution of inter-skyrmion correlations.}

Our work therefore identifies topological-defect correlations as the key
degree of freedom underlying the dielectric response and, more
generally, as an organizing principle for functional responses in ferroic
materials. The dielectric response is governed not only by local polar
distortions but also by the collective organization of the skyrmion ensemble
(see Sec.~SVI). Such collective correlations are also expected to
influence a wide range of functional properties, including dielectric,
nonlinear-optical~\cite{74d5-4hsw}, and thermal-transport
responses, thereby providing a potential strategy for
engineering the functional and dynamical properties of ferroic materials
through controlled correlations of topological defects.

{\it Acknowledgments---}This work is supported by the Army Research Office under the ETHOS MURI via cooperative agreement W911NF-21-2-0162 for the development of oxide superlattice materials.\\

All the simulations in this work are done by the commercial software MuPRO. Long-Qing Chen has a financial interest in MuPRO, LLC, a company which licenses and markets the software package used in this research.\\

\bibliography{reference.bib}

\end{document}


\title{Supplemental Materials: Hidden Crossover and Relaxor-Like Response from Emerging Polar Skyrmion Correlations in Ferroelectric Superlattices}

\author{Zhi-Yang Wang}

\affiliation{Department of Materials Science and Engineering and Materials Research Institute, The Pennsylvania State University, University Park, PA 16802, USA}

\author{Fei Yang}
\email{fzy5099@psu.edu}

\affiliation{Department of Materials Science and Engineering and Materials Research Institute, The Pennsylvania State University, University Park, PA 16802, USA}

\author{Long-Qing Chen}
\email{lqc3@psu.edu}

\affiliation{Department of Materials Science and Engineering and Materials Research Institute, The Pennsylvania State University, University Park, PA 16802, USA}

\date{\today}

\maketitle  

\tableofcontents

\section{Model and material coefficients}
To simulate the evolution of polar topological phases in 
Pb$_x$Sr$_{1-x}$TiO$_3$/PbTiO$_3$ superlattices, the polarization 
$\mathbf{P} = (P_1, P_2, P_3)$ is taken as the order parameter. The time 
evolution of $\mathbf{P}$ is governed by the time-dependent 
Ginzburg-Landau (TDGL) equation~\cite{ref1}:
\begin{equation}
    \frac{\partial \mathbf{P}(\mathbf{r},t)}{\partial t} 
    = -\Gamma \frac{\delta F}{\delta \mathbf{P}(\mathbf{r},t)},
\end{equation}
where $\mathbf{r}$ and $t$ denote the spatial coordinate and time, 
respectively, and $\Gamma$ is the kinetic coefficient.

The total free energy of the system is written as
\begin{equation}
    F = \int \left( f_{\text{Landau}} + f_{\text{grad}} 
    + f_{\text{elec}} + f_{\text{elas}} \right)\, dV,
\end{equation}
where $f_{\text{Landau}}$, $f_{\text{grad}}$, $f_{\text{elec}}$, and 
$f_{\text{elas}}$ denote the Landau, gradient, electrostatic, and elastic 
energy densities, respectively, as described in the main text.

The Poisson equation and the mechanical equilibrium equation are solved 
self-consistently with the TDGL equation to obtain the evolution of the 
electric potential, elastic displacement, and polarization 
field~\cite{ref2}.

The Landau coefficients ($a_i$, $a_{ij}$, $a_{ijk}$) and the 
electrostrictive coefficients ($Q_{ij}$) are taken from 
Ref.~\cite{ref3}, while the elastic compliances ($S_{ij}$) are taken from 
Ref.~\cite{shirokov2022thermodynamic}. All material parameters are 
summarized in Table~\ref{para}.

\begin{table}[t]
\centering
\label{parameterTable}
\caption{Material coefficients of Pb$_x$Sr$_{1-x}$TiO$_3$~\cite{ref3,shirokov2022thermodynamic}.}
\renewcommand{\arraystretch}{1.4}
\begin{tabular}{ll}
\toprule
\textbf{Coefficient} & \textbf{Expression / Value} \\
\hline
$a_1$ ($\mathrm{C^{-2} m^2 N}$) 
& $\dfrac{T_s}{2\epsilon_0 C}
\left[\coth\!\left(\dfrac{T_s}{T}\right) 
- \coth\!\left(\dfrac{T_s}{T_c}\right)\right]$ \\

$C$ ($10^5\,\mathrm{K^{-1}}$) 
& $1.50 - 0.995(1-x)$ \\

$T_c$ (K) 
& $752 - 697(1-x)$ \\

$T_s$ (K) 
& $54$ \\

$a_{11}$ ($\mathrm{C^{-4} m^6 N}$) 
& $\left[17.73(1-x) - 0.73\right]\times 10^8$ \\

$a_{12}$ ($\mathrm{C^{-4} m^6 N}$) 
& $\left[24(1-x)^{3/2} + 7.5\right]\times 10^8$ \\

$a_{111}$ ($\mathrm{C^{-6} m^{10} N}$) 
& $\left[20(1-x) + 2.6x\right]\times 10^8$ \\

$a_{112}$ ($\mathrm{C^{-6} m^{10} N}$) 
& $\left[8(1-x) + 6.1x\right]\times 10^8$ \\

$a_{123}$ ($\mathrm{C^{-6} m^{10} N}$) 
& $\left[20(1-x) - 3.7x\right]\times 10^9$ \\

$Q_{11}$ ($10^{-2}\,\mathrm{m^4 C^{-2}}$) 
& $8.9 - 3.81(1-x)$ \\

$Q_{12}$ ($10^{-2}\,\mathrm{m^4 C^{-2}}$) 
& $-1.31(1-x) - 2.6x$ \\

$Q_{44}$ ($10^{-2}\,\mathrm{m^4 C^{-2}}$) 
& $1.065(1-x) + 6.75x$ \\

$S_{11}$ ($\mathrm{m^2/N}$) 
& $8.0\times 10^{-12}x + 3.52\times 10^{-12}(1-x)$ \\

$S_{12}$ ($\mathrm{m^2/N}$) 
& $-2.5\times 10^{-12}x - 0.85\times 10^{-12}(1-x)$ \\

$S_{44}$ ($\mathrm{m^2/N}$) 
& $9.0\times 10^{-12}x + 7.87\times 10^{-12}(1-x)$ \\

$g$ ($C^{-2}m^4N$)
& $1.0211\times 10^{-10}$ \\

\hline
\hline
\end{tabular}
\label{para}
\end{table}

\section{Boundary conditions}
In the $x$-$y$ plane, periodic boundary conditions are applied to the 
TDGL equation, the Poisson equation, and the mechanical equilibrium equation. 
Along the $z$ direction, a homogeneous Neumann boundary condition is imposed 
for the TDGL equation:
\begin{equation}
    \left. \frac{\partial P_i}{\partial z} \right|_{z=0} = 
    \left. \frac{\partial P_i}{\partial z} \right|_{z=h_f} = 0.
\end{equation}
Here, $h_f$ denotes the thickness of the superlattice, and $z=0$ corresponds 
to the substrate-superlattice interface.

For the mechanical equilibrium equation, thin-film elastic boundary conditions 
are used along the $z$ direction. The top surface of the superlattice is 
stress-free, while the displacement is fixed at the bottom of the substrate~\cite{ref6}:
\begin{equation}
    \left. \sigma_{i3} \right|_{z=h_f} = 0, \qquad 
    \left. u_i \right|_{z=-h_s} = 0.
\end{equation}
Here, $h_s$ is the thickness of the substrate.

For the Poisson equation, both short-circuit and open-circuit boundary 
conditions are considered in the simulations. Under short-circuit boundary conditions, the electric potential is fixed 
at the top surface of the superlattice, while the potential at the 
substrate-superlattice interface is determined by the applied external 
electric field:
\begin{equation}
    \left. \phi \right|_{z=h_f} = 0, \qquad 
    \left. \phi \right|_{z=0} = \phi_{\text{ext}}.
\end{equation}

Under open-circuit boundary conditions, the out-of-plane electric 
displacement is set to zero at both the top surface and the 
superlattice-substrate interface~\cite{ref7}:
\begin{equation}
    \left. D_3 \right|_{z=h_f} = 0, \qquad 
    \left. D_3 \right|_{z=0} = 0.
\end{equation}

\section{Simulation setups}
To avoid round-off errors, the distance $r$ and time $t$ are rescaled into 
dimensionless variables as $r^* = r / r_0$, $\tau = t / \tau_0$, and 
$p_i = P_i / P_0$. In terms of these normalized variables, the effective 
coefficients are given by
\begin{equation}
\begin{split}
    a_i^* &= \frac{a_i}{a_0}, \qquad
    a_{ij}^* = \frac{a_{ij} P_0^2}{a_0}, \qquad
    a_{ijk}^* = \frac{a_{ijk} P_0^4}{a_0},\\ 
    g^* &= \frac{g}{a_0 r_0^2}, \qquad
    Q_{ij}^*= Q_{ij} P_0^2, \qquad
    S_{ij}^*= S_{ij} a_0 P_0^2 .
\end{split}
\end{equation}
Here we choose $r_0 = 10^{-9}\,\mathrm{m}$, 
$P_0 = 0.807\,\mathrm{C/m^2}$ (the spontaneous polarization of PTO at $298\,\mathrm{K}$, calculated from the coefficients in Table~\ref{para}), 
and $\tau_0 = 1/(a_0 \Gamma)$ with $a_0 = a_1(x=1, T = 298\,\mathrm{K})$, 
such that in reduced units $a_1^*(x=1, T = 298\,\mathrm{K}) = -1$. 
The reduced parameter $g^*$ is set to $0.6$, and the background dielectric constant is taken as $\kappa_b = 40$ for all simulations. {It is noted that the time-dependent Ginzburg-Landau equation is expressed in terms of the normalized time variable $\tau=t/\tau_0=ta_0\Gamma$. As a result, the kinetic coefficient $\Gamma$ is absorbed into the time scaling and does not appear as an independent parameter in the dimensionless evolution equation.}

A computational grid of $200\Delta \times 200\Delta \times 200\Delta$ 
is employed, with $\Delta = 0.4$. Along the $z$ direction, the substrate, 
superlattice, and vacuum regions occupy 20, 176, and 4 grid points, respectively.  The superlattice consists of 16 unit-cell layers of 
Pb$_x$Sr$_{1-x}$TiO$_3$ grown on a (001)-oriented SrTiO$_3$ substrate, 
followed by 16 layers of PbTiO$_3$ and 16 layers of 
Pb$_x$Sr$_{1-x}$TiO$_3$, forming a periodic structure repeated five times. We have verified that further increasing the grid resolution does not qualitatively change the polarization patterns or susceptibility results. 
The spatial variation of material properties, including the Landau 
coefficients, elastic stiffness, electrostrictive coefficients, and 
reference lattice parameters, is used to define the superlattice. The film-substrate interface is assumed to be coherent, such that the 
lattice mismatch strain between the SrTiO$_3$ (001) substrate and the 
PbTiO$_3$/Pb$_x$Sr$_{1-x}$TiO$_3$ layers is determined by their lattice 
parameters. The pseudo-cubic lattice parameter of 
Pb$_x$Sr$_{1-x}$TiO$_3$ at room temperature is fitted from 
Ref.~\cite{ref5} as
\begin{equation}
    c_x = 0.03797(1-x)^2 - 0.08878(1-x) + 3.9567.
\end{equation}
In principle, thermal expansion modifies the lattice mismatch as the 
temperature increases. However, experimental measurements of the thermal 
expansion coefficient for Pb$_x$Sr$_{1-x}$TiO$_3$~\cite{ref5} indicate 
that this variation is two orders of magnitude smaller than the lattice 
mismatch at room temperature. Therefore, the effect of thermal expansion 
is neglected in the present work.

In this work, the time-dependent TDGL equation is 
solved using a semi-implicit Fourier spectral method, while the Poisson 
equation and the mechanical equilibrium equation are solved using a 
Fourier spectral method. The numerical integration is performed using a dimensionless time step $\Delta\tau=0.02$. Details of the numerical implementation, 
including spectral techniques, boundary conditions, and elastic inhomogeneity, can be found in 
Refs.~\cite{chen1998applications,ref6,ref7,ref11,ref12}. To mimic the growth process of the superlattice, open-circuit boundary 
conditions are imposed before electrode deposition. Specifically, the 
system is first evolved for $10^4$ time steps under open-circuit 
conditions, followed by an additional $5\times 10^3$ steps under 
short-circuit boundary conditions.

{The initial skyrmion configurations are generated by evolving random polarization fields at 800 K. Specifically, the initial in-plane polarization components are sampled from a uniform distribution in the interval $[-0.005,0.005]$. For the out-of-plane component, structural asymmetries commonly present in experiments, such as asymmetric electrodes and flexoelectric fields arising from intrinsic strain gradients~\cite{ref8,ref9,ref10}, may bias the polarization evolution. To incorporate this effect in a simplified manner, the initial out-of-plane polarization is sampled from a shifted uniform distribution, $[-0.006,0.004]$. The system is then further evolved for an additional $10^4$ time steps to ensure equilibration. Subsequently, the temperature is gradually reduced from 800 K to 50 K in steps of 50 K. At each temperature, the final configuration obtained at the preceding temperature is used as the initial condition, and the system is evolved for another $10^4$ time steps before physical quantities, such as the dielectric susceptibility, are evaluated. Consequently, all configurations shown in the main text are obtained along the same continuous cooling trajectory rather than from independent simulations at each temperature. The emergence of interlayer-correlated skyrmion states upon cooling therefore reflects the intrinsic temperature-driven evolution of the system along a well-defined thermodynamic path.}

\begin{figure}[h]
    \centering
    \includegraphics[width=1\linewidth]{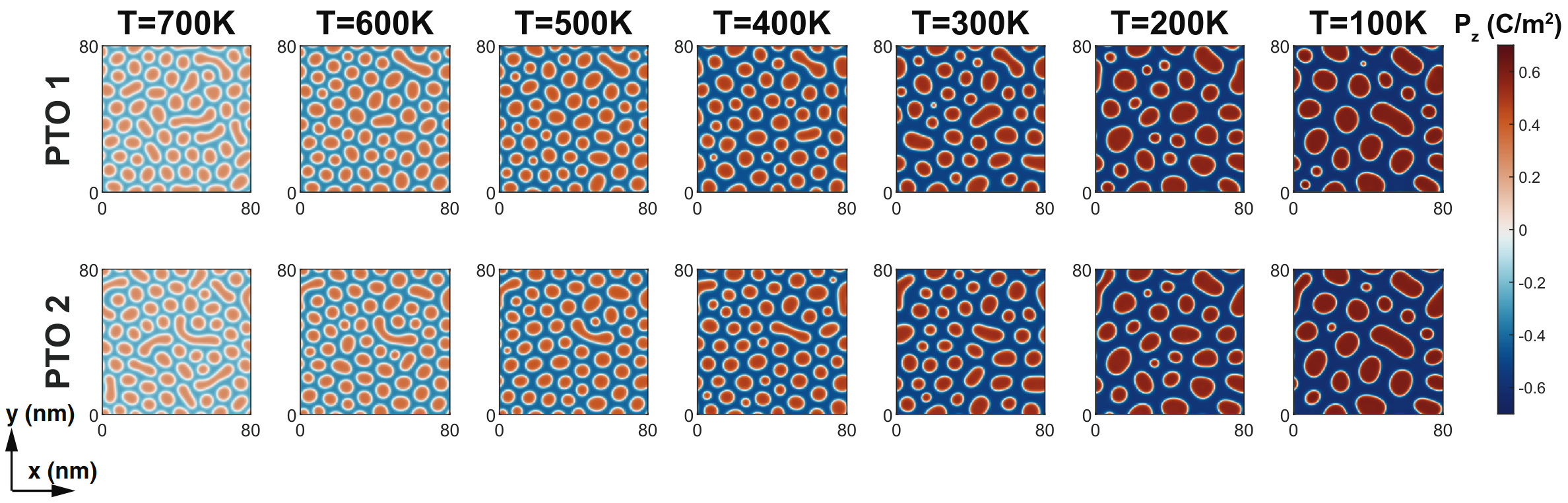}
    \caption{{Evolution of the skyrmion configurations during continuous cooling for two adjacent PTO layers in the superlattice. The configurations shown at each temperature are obtained sequentially during the same cooling process. }}
    \label{fig:Cooling_xy}
\end{figure}

{Figure~\ref{fig:Cooling_xy} shows the complete cooling protocol and the corresponding evolution of the skyrmion configurations during continuous cooling. As shown, at high temperatures, the skyrmion patterns in different layers exhibit little interlayer correlation; for example, at 700K or 600 K, each layer displays a distinct skyrmion configuration. As the temperature decreases, the degree of interlayer correlation progressively increases. In the low-temperature regime (e.g., 200~K or 100~K), the skyrmion configurations become strongly correlated across different layers, with highly similar patterns emerging throughout the superlattice, indicating the establishment of interlayer correlations.} 

A slow drift of the net out-of-plane polarization persists throughout the simulations. To eliminate the influence of this drift, all susceptibilities are extracted from two parallel simulations performed under identical conditions, one with an external electric field and the other without. For the DC susceptibility, the driven simulation is evolved for $10^3$ time steps under a bottom bias $\phi_{\rm bot}=7.04\times10^{-3}\,\mathrm{V}$, corresponding to an electric field $E=10^5\,\mathrm{V/m}$. The resulting polarization change is denoted by $\Delta P_E$. In the reference simulation, the system evolves for the same duration without an applied field, yielding a polarization change $\Delta P_{\rm ref}$. The net field-induced response is obtained as
\begin{equation}
\Delta P_{\rm net}=\Delta P_E-\Delta P_{\rm ref},
\end{equation}
from which the static susceptibility is calculated as
\begin{equation}
\chi=\frac{\Delta P_{\rm net}}
{\epsilon_0 E}.
\end{equation}
For the AC susceptibility, the same subtraction procedure is employed. In the driven simulation, a sinusoidal bias $\phi_{\rm bot}=7.04\times10^{-3}\sin(2\pi\tau/t_p)\,\mathrm{V}$ is applied, where
\begin{equation}
t_p=nt_0\ (n=1,2,3,4),\quad t_0=1000\Delta\tau,\quad \Delta\tau=0.01.
\end{equation}
The reference simulation is performed without a field over the same duration. The net response is
$\Delta P_{\rm net}(\tau)=\Delta P_E(\tau)-\Delta P_{\rm ref}(\tau)$, which is fitted as
\begin{equation}
\Delta P_{\rm net}(\tau)=\Delta P_0\sin(2\pi\tau/t_p-\delta)+C.
\end{equation}
The real and imaginary susceptibilities are then
\begin{equation}
\chi'=\frac{\Delta P_0}{\epsilon_0 E}\sin\delta,\qquad
\chi''=\frac{\Delta P_0}{\epsilon_0 E}\cos\delta.
\end{equation}

We emphasize that the present study relies on large-scale, fully 
three-dimensional phase-field simulations on a $200\times200\times200$ 
grid, with self-consistent solutions of the TDGL, Poisson, and mechanical 
equilibrium equations at each time step. The substantial computational 
cost associated with the long-time evolution ($\sim 10^4$-$10^5$ steps) 
and multiple parallel simulations for susceptibility extraction ensures 
numerical convergence and robustness of the results. These large-scale simulations enable us to resolve the emergence and 
evolution of polar topological structures over mesoscopic length and 
time scales, which are essential for capturing the hidden crossover 
and relaxor-like response reported in this work.

\section{Polar skyrmions and polar vortices}

Polar skyrmions are topologically nontrivial polarization textures formed by
the continuous rotation of the local polarization vector in real space~\cite{junquera_topological_2023-2}. In a
ferroelectric system, the polarization field
$\mathbf{P}(\mathbf{r})=(P_1,P_2,P_3)$ serves as the order parameter. A
skyrmion-like texture is characterized not simply by a spatial modulation of
the magnitude of $\mathbf{P}$, but by a nontrivial winding of its direction.  In a Landau-type description, the
polarization order parameter may be decomposed schematically into an
amplitude and an orientation,
\begin{equation}
    \mathbf{P}(\mathbf{r},t)
    =
    P(\mathbf{r},t)\,\mathbf{n}(\mathbf{r},t),
\end{equation}
where $P=|\mathbf{P}|$ describes the amplitude and $\mathbf{n}$ describes
the local direction.  Amplitude fluctuations correspond primarily to
variations of $P(\mathbf{r},t)$, namely local softening or enhancement of the
polarization magnitude. In analogy with broken-symmetry systems, the
oscillation of the order-parameter magnitude is often referred to as an
amplitude mode or Higgs-like mode~\cite{shimano2020higgs,yang2020theory,matsunaga2013higgs,matsunaga2014light,yang2024optical,yang2023optical,PhysRevB.106.L081105}.  
 By contrast, a polar skyrmion is mainly a spatially organized orientational
texture of $\mathbf{n}(\mathbf{r})$. Its essential feature is the winding of
the polarization direction. In a two-dimensional plane perpendicular to the skyrmion tube, the
topological charge density is given by~\cite{junquera_topological_2023-2,Nagaosa2013}
\begin{equation}\label{charge}
    q(\mathbf{r})=
    \frac{1}{4\pi}
    \mathbf{n}\cdot
    \left(
    \frac{\partial \mathbf{n}}{\partial x}
    \times
    \frac{\partial \mathbf{n}}{\partial y}
    \right),
\end{equation}
and the corresponding skyrmion number is
\begin{equation}\label{number}
    N_{\rm sk}=\int q(\mathbf{r})\,d^2r .
\end{equation}
For an ideal isolated skyrmion, $N_{\rm sk}$ is close to an integer,
typically $N_{\rm sk}=\pm 1$, depending on the winding chirality. In the
present three-dimensional simulations, such quantities can be evaluated
layer by layer along the out-of-plane direction to identify skyrmion tubes
and their spatial correlations. The formation of polar skyrmions originates from the competition among the
Landau, gradient, electrostatic, and elastic contributions to the total free
energy. The gradient energy penalizes sharp spatial variations of
polarization and favors smooth rotations, while the electrostatic energy
promotes flux-closure configurations to reduce depolarization fields. The
elastic energy, through electrostrictive coupling and epitaxial constraint,
further favors spatially modulated polarization structures. As a result,
polar skyrmions may emerge as mesoscale textures stabilized by the
self-consistent coupling among polarization, electric potential, and elastic
strain.\\

{It should be noted that polar vortices and polar skyrmions represent two distinct topological polarization textures. A polar vortex is characterized by a
nonzero winding number of the polarization field,
\begin{equation}
W=\frac{1}{2\pi}\oint_C \nabla\theta(\mathbf{r})\cdot d\mathbf{l},
\end{equation}
where $\theta(\mathbf{r})$ denotes the local polarization angle in the plane
of polarization rotation and $C$ is a closed contour surrounding the vortex
core.  For an isolated vortex, $W=\pm1$, with the sign determined by the
sense of polarization rotation. Typically, in ferroelectric/dielectric superlattices, vortices and stripe domains are related: the
stripe domains carry polarization oriented primarily along the stacking direction of the superlattice, and
the vortex structures appear at the domain walls between adjacent stripes, where the dipoles rotate within
a plane that is perpendicular to the ferroelectric-dielectric interfaces and parallel to the stacking direction.
The in-plane polarization component is thus localized at the domain walls.   A polar skyrmion, on the other hand, is
characterized by the two-dimensional wrapping of the polarization field
[Eqs.~(\ref{charge}) and~(\ref{number}); see also
Fig.~\ref{fig:Cooling_xy}], with the polarization continuously evolving
from one out-of-plane orientation near the skyrmion core toward the
opposite orientation outside the core, accompanied by an intermediate
in-plane rotation.
}

{
The relative stability of these different polarization textures is
determined by the competition among the elastic, electrostatic, gradient,
and Landau contributions to the free energy and is known to be highly
sensitive to epitaxial strain and superlattice geometry. In particular, the
elastic contribution depends on the full spatial polarization texture
through the spontaneous strain. Different vortex,
stripe, and skyrmion textures produce different spatial distributions of
spontaneous strain and consequently different elastic-energy costs under
the epitaxial constraint. 
This energetic selection is directly demonstrated by  previous studies
of PTO/STO superlattices~\cite{wang_tuning_2024,10.1063/5.0160901}. By
changing the epitaxial strain, while retaining the same phase-field
framework, the system can be driven from a skyrmion regime into a vortex
regime. Similarly, even on the same STO substrate, reducing the PTO-layer
thickness can stabilize a vortex phase, whereas thicker PTO layers favor
polar skyrmions. These results demonstrate that vortex states are fully
accessible within the present phase-field framework and that their absence
under the conditions considered here 
results from the specific energetic balance associated with the present epitaxial
constraint and superlattice geometry. 
For the systems considered in the present work, the available experimental
evidence and our phase-field simulations consistently indicate that the
polar-skyrmion state is favored under the relevant STO-substrate constraint
and layer geometry. Previous experimental studies of PTO/STO superlattices
grown on STO substrates~\cite{das_observation_2019,das_local_2021-1,
Gong2023,Li2025,mccarter_structural_2022} have reported polar skyrmion
textures under comparable conditions. Moreover, our simulations of
Pb$_x$Sr$_{1-x}$TiO$_3$/PTO superlattices are initialized from random
disordered polarization configurations at high temperature and subsequently
evolved through a continuous cooling procedure without imposing any
specific topological texture. Nevertheless, the system consistently
develops skyrmion configurations upon cooling. These
results indicate that the skyrmion texture is energetically favored over
the stripe/vortex texture within the temperature, composition, strain, and
geometrical regime investigated in the present work.
}

\section{Interlayer correlations of skymion}

\subsection{Coupling between the adjacent PTO layers}

\begin{figure}
    \centering
    \includegraphics[width=\linewidth]{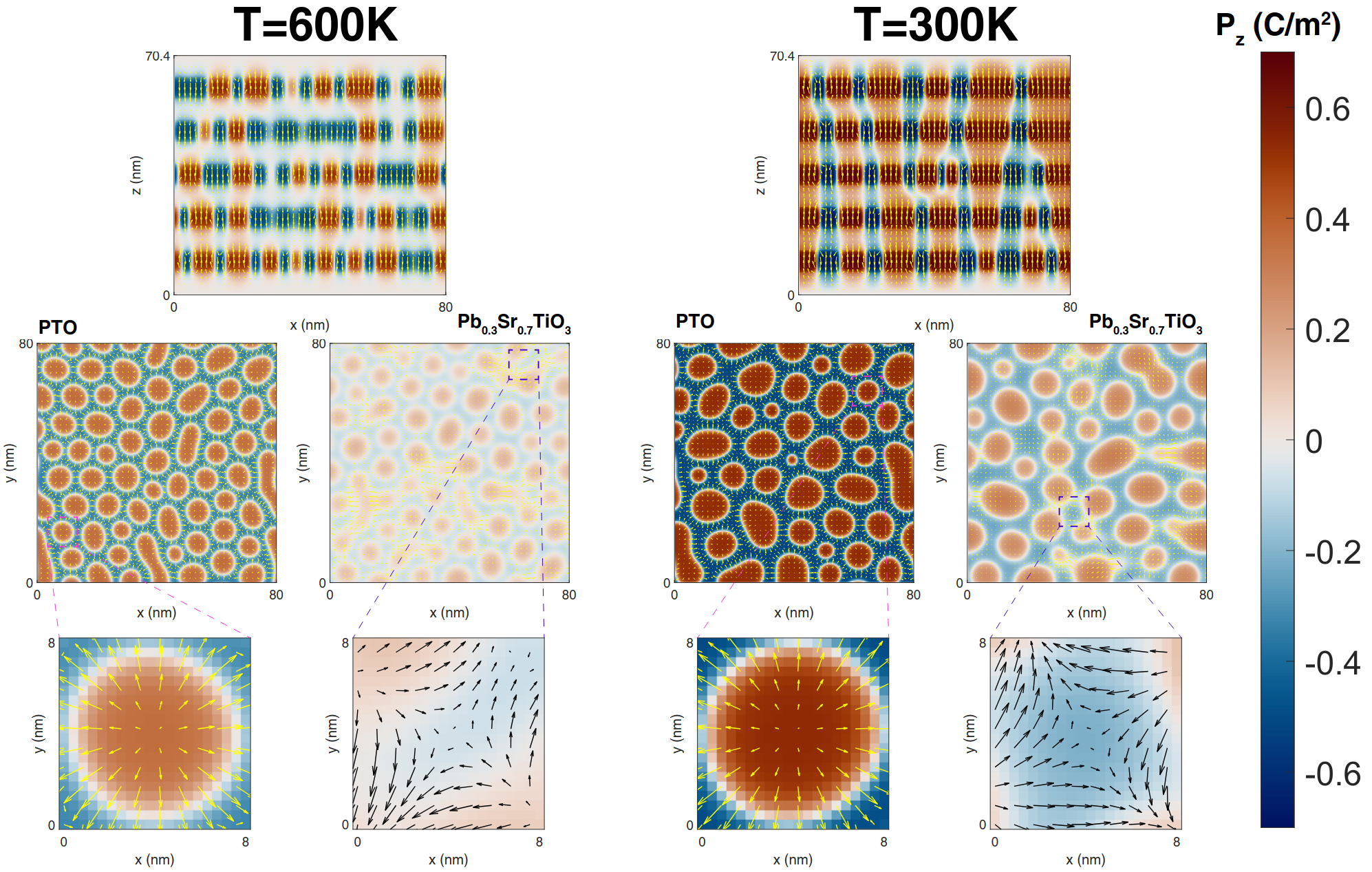}
    \caption{{Polarization textures of the Pb$_{0.3}$Sr$_{0.7}$TiO$_3$/PTO superlattice at 300 K and 600 K. The top row shows vertical cross sections of the polarization field in the $x$-$z$ plane, and the middle row shows horizontal cross sections of representative PTO layer and Pb$_{0.3}$Sr$_{0.7}$TiO$_3$ layer. The bottom row illustrates representative skyrmion and antiskyrmion polarization textures identified in the superlattice. Colors represent the out-of-plane polarization component, $P_z$, whereas arrows indicate the polarization vectors projected onto the corresponding plane.}}
    \label{fig:topo}
\end{figure}

{Figure~\ref{fig:topo} shows the detailed polarization-field distributions at representative temperatures. 
As can be seen from the $x$-$z$ cross-sectional views, the Pb$_x$Sr$_{1-x}$TiO$_3$ spacer layers exhibit only weak polarization at high temperatures. As a result, dipolar interactions are transmitted only weakly through the spacer, effectively decoupling neighboring PTO layers and leading to largely uncorrelated skyrmion configurations in different layers. Upon cooling, however, the polarization of the  Pb$_x$Sr$_{1-x}$TiO$_3$ layers increases substantially. The enhanced polarization within the spacer facilitates the propagation of dipolar interactions across the superlattice, thereby strengthening the coupling between skyrmions in adjacent PTO layers. Consequently, the skyrmion configurations evolve from weakly correlated arrangements at high temperatures to strongly correlated states at low temperatures.} 

{The polarization maps therefore provide direct real-space evidence supporting our interpretation that the development of interlayer skyrmion correlations is driven by the increasing polarizability of the dielectric spacer layers upon cooling.}\\

{It should be note that the antiskyrmion-like polarization textures are also present in our simulations, as highlighted by the dashed red squares in Fig.~\ref{fig:topo}. In particular, the local polarization configurations exhibit features analogous to the antiskyrmionic textures reported in Ref.~\cite{Abid2021}, including a reversal of the polarization winding pattern relative to the surrounding skyrmion textures. While the primary focus of the present work is the emergence of interlayer skyrmion correlations, the coexistence of such antiskyrmion-like defects further illustrates the rich topological landscape of the polar textures realized in these superlattices.}

\subsection{Robust feature in the low-temperature regime}

\begin{figure}[h]
    \centering
    \includegraphics[width=0.9\linewidth]{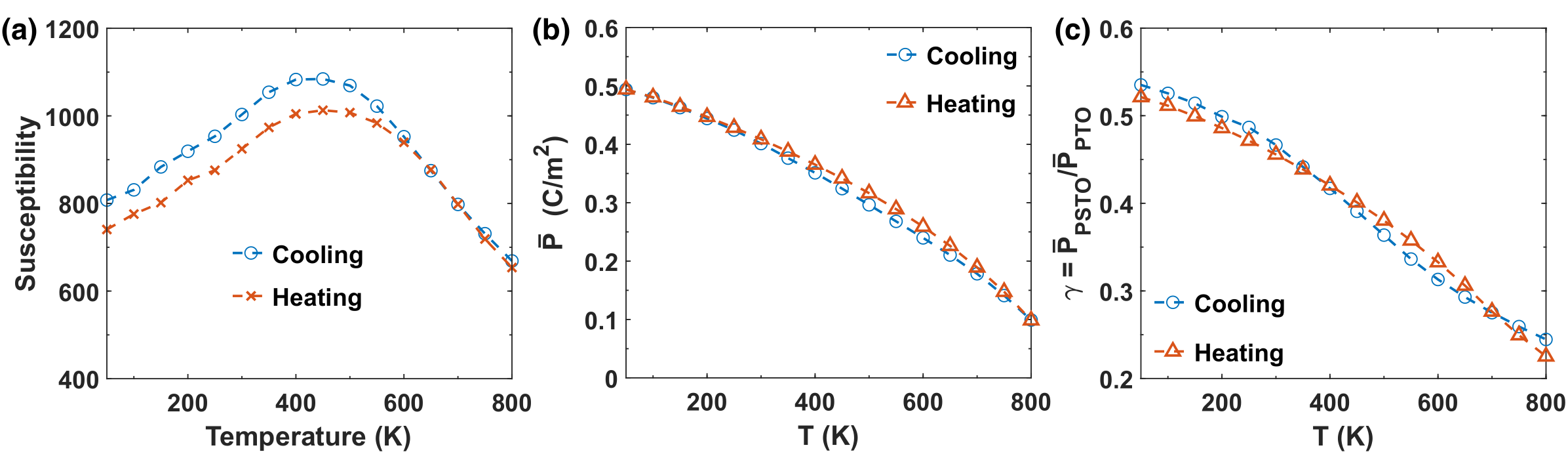}
\caption{{Temperature dependence of ({\bf a}) DC dielectric susceptibility,  ({\bf b}) the spatially averaged polarization magnitude $\overline{P}(T)$, and ({\bf c}) the polarization ratio $\gamma(T)$ between the Pb$_x$Sr$_{1-x}$TiO$_3$ and PTO layers, obtained during cooling and heating cycles.}}
    \label{fig:tcycle}
\end{figure}

{The resulting trajectory obtained from the heating simulations is shown in Fig.~\ref{fig:tcycle} and compared with that from the cooling simulations. The heating and cooling paths have only a minor influence on the overall temperature dependence of the polarization [Fig.~\ref{fig:tcycle}(b)] within the ferroelectric phase $(T<T_c)$. They lead to slightly different polarization ratios $\gamma(T)$ between the Pb$_x$Sr$_{1-x}$TiO$_3$ and PTO layers [Fig.~\ref{fig:tcycle}(c)]. As the development of interlayer skyrmion correlations is sensitive to the partitioning of polarization between the constituent layers, these differences modify the temperature evolution of the interlayer correlations. Consequently, the crossover temperature associated with the correlation buildup, and thus the temperature of the dielectric peak [Fig.~\ref{fig:tcycle}(a)], exhibits a modest shift between the heating and cooling trajectories.}

\begin{figure}[h]
    \centering
    \includegraphics[width=0.6\linewidth]{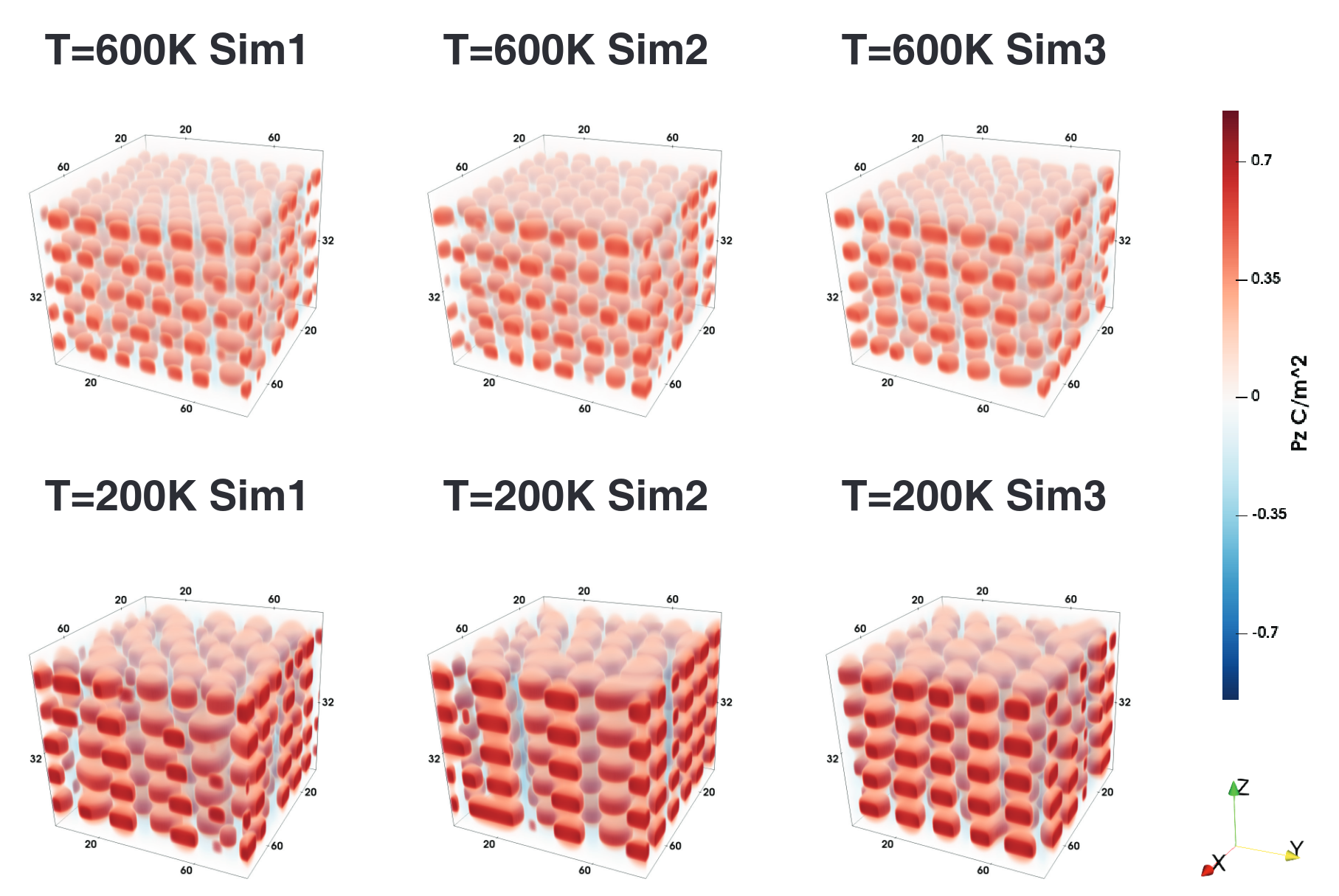}
    \caption{{Simulation results obtained by evolving three independent disordered initial configurations at 200 K and 600 K to their respective equilibrium states.}}
    \label{fig:200K600K}
\end{figure}

{The primary objective of the present work is to demonstrate the emergence and development of interlayer skyrmion correlations in the low-temperature regime. The relevant thermodynamic anomaly is governed primarily by the existence and evolution of these interlayer correlations. While the apparent crossover temperature exhibits a modest shift between the cooling and heating trajectories, the interlayer correlations strengthen persist throughout the low-temperature regime.
 To demonstrate this point, we perform additional simulations in which the system was initialized from independent disordered configurations at several selected temperatures. As shown in Fig.~\ref{fig:200K600K}, we find that, while different initial conditions can lead to variations in the detailed intralayer skyrmion arrangements, robust interlayer correlations are established in all cases in the low-temperature regime. The reproducibility of the interlayer correlation from different disordered initial states suggests that the correlated state possesses a large basin of attraction in the low-temperature energy landscape,  and indicates that the interlayer correlation identified in our work is an intrinsic low-temperature property of the system.} 

\section{Effective analytical model}

{In this part, we present the theoretical analysis of the dielectric response. By deriving both the static and frequency-dependent susceptibilities while explicitly incorporating the contribution of interlayer skyrmion correlations, we establish an explicit   connection between the evolution of the skyrmion texture and the observed dielectric anomaly.}

\subsection{Static dielectric susceptibility}

{We first derives an effective analytical model to illustrate how the spatial organization of polar skyrmions renormalizes the dielectric susceptibility. The analytical result well captures the temperature dependence of the dielectric susceptibility obtained from the full phase-field simulations. This supports our physical interpretation of the susceptibility peak.}

{Specifically, we write the polarization field in the $l$-th ferroelectric layer as
\begin{equation}
    {\bf P}_{l}({\bf r})
    =
    P({\bf r})\,{\bf n}_{l}({\bf r}),
    \qquad
   {\bf r}=(x,y),
    \label{eq:P_amplitude_texture}
\end{equation}
where $P({\bf r})$ denotes the characteristic polarization amplitude and
${\bf n}_{l}({\bf r})$ is a dimensionless field describing the local
polarization orientation and the skyrmion texture, with $|{\bf n}_{l}({\bf r})|=1$. Equation~\eqref{eq:P_amplitude_texture} separates the
relatively uniform amplitude degree of freedom from the spatially
nonuniform orientational texture. The spatially averaged out-of-plane polarization is
\begin{equation}
    \overline{P}_{z}
    =
    \frac{1}{N\mathcal A}
    \sum_{l=1}^{N}
    \int_{\mathcal A}d^{2}r\,
    P n_{l,z}({\bf r})
    =
    m_z P,
    \label{eq:Pbarz_definition}
\end{equation}
where $
    m_z
    =
    \frac{1}{N\mathcal A}
    \sum_{l=1}^{N}
    \int_{\mathcal A}d^{2}r\,
    n_{l,z}({\bf r})$ is the geometrical projection factor of the polarization  texture onto the
direction of the applied electric field. Here $N$ is the number of
ferroelectric layers and $\mathcal A$ is the in-plane area. For a spatially uniform electric field $E_z$, the field-coupling energy
density is 
\begin{equation}
    f_E=-E_z\overline{P}_z=-m_zE_zP.
    \label{eq:field_coupling_texture}
\end{equation}
\begin{itemize}
\item For the polarization
amplitude $P$, one can use a scalar Landau expansion,
\begin{equation}
    f_{\rm L}(P)
    =
    \frac{\alpha}{2}P^2
    +
    \frac{\beta}{4}P^4
    +
    \frac{\gamma}{6}P^6.
    \label{eq:Landau_scalar}
\end{equation}
The temperature dependence is mainly contained in $\alpha(T)$, while
$\beta$ and $\gamma$ are taken to be weakly $T$-dependent. 
The local Landau curvature evaluated at an equilibrium polarization
amplitude $P_0$ is
\begin{equation}
    K(T)
    =
    \left.
    \frac{\partial^2f_{\rm L}}{\partial P^2}
    \right|_{P=P_0}
    =
    \alpha
    +
    3\beta P_0^2
    +
    5\gamma P_0^4.
    \label{eq:Landau_curvature}
\end{equation}
\end{itemize}}

{The skyrmion texture generates additional quadratic contributions to
the free energy through both the in-plane and out-of-plane gradient
energies. The in-plane gradient energy is written as
\begin{equation}
    F_{g,\perp}
    =
    \frac{\kappa_\perp}{2}
    \sum_l
    \int_{\mathcal A}d^2r
    \left[
    |\partial_x{\bf P}_l|^2
    +
    |\partial_y{\bf P}_l|^2
    \right],
    \label{eq:F_gradient_inplane}
\end{equation}
where $\kappa_\perp$ is the in-plane polarization-gradient coefficient.
Using Eq.~\eqref{eq:P_amplitude_texture}, one obtains
\begin{equation}
    F_{g,\perp}
    =
    \frac{\kappa_\perp P^2}{2}
    \sum_l
    \int_{\mathcal A}d^2r
    |\nabla_\perp{\bf n}_l|^2.
    \label{eq:F_gradient_inplane_texture}
\end{equation}}

{We define the normalized in-plane skyrmion wave-vector scale as
\begin{equation}
    q_{\rm sk}^2
    =
    \frac{
    \displaystyle
    \sum_l\int_{\mathcal A}d^2r
    |\nabla_\perp{\bf n}_l|^2
    }{
    \displaystyle
    \sum_l\int_{\mathcal A}d^2r
    |{\bf n}_l|^2
    }=\frac{1}{N\mathcal A}
    \displaystyle
    \sum_l\int_{\mathcal A}d^2r
    |\nabla_\perp{\bf n}_l|^2
    .
    \label{eq:qsk_realspace}
\end{equation}
The in-plane gradient energy density can then be expressed as
\begin{equation}
    f_{g,\perp}
    =
    \frac{1}{2}
    \kappa_\perp q_{\rm sk}^2P^2.
    \label{eq:fg_inplane_effective}
\end{equation}
Consequently, the internal skyrmion texture produces the quadratic
renormalization
\begin{equation}    
\Delta\alpha_{\rm sk}=\kappa_\perp q_{\rm sk}^2.\label{eq:alpha_skyrmion}
\end{equation}}

{Because the polar skyrmions reside in periodically separated
ferroelectric layers, their interlayer coupling is more naturally
described by a discrete gradient energy. Let $d$ denote the distance
between two equivalent neighboring ferroelectric layers or neighboring
PTO blocks. The out-of-plane gradient energy is
\begin{equation}
    F_{g,z}
    =
    \frac{\kappa_z}{2d^2}
    \sum_l
    \int_{\mathcal A}d^2r
    \left|
    {\bf P}_{l+1}({\bf r})
    -
    {\bf P}_l({\bf r})
    \right|^2.
    \label{eq:F_gradient_z_discrete}
\end{equation}
Substituting Eq.~\eqref{eq:P_amplitude_texture} gives
\begin{equation}
    F_{g,z}
    =
    \frac{\kappa_zP^2}{2d^2}
    \sum_l
    \int_{\mathcal A}d^2r
    \left|
    {\bf n}_{l+1}({\bf r})
    -
    {\bf n}_l({\bf r})
    \right|^2.
    \label{eq:F_gradient_z_texture}
\end{equation}
Using $|{\bf n}_{l+1}-{\bf n}_l|^2
    =|{\bf n}_{l+1}|^2+|{\bf n}_l|^2-2{\bf n}_{l+1}\cdot{\bf n}_l$ and $|{\bf n}_l|=1$, we obtain $|{\bf n}_{l+1}-{\bf n}_l|^2=2(1-{\bf n}_{l+1}\cdot{\bf n}_l)$.}

{We define the interlayer structural correlation function as 
\begin{equation} 
\mathcal C(md) = \frac{1}{N\mathcal A} \sum_l \int_{\mathcal A}d^2r\,{\bf P}_{l+m}({\bf r}) \cdot {\bf P}_{l}({\bf r}).\label{eq:C_discrete_definition} \end{equation} 
The average mismatch between neighboring layers is then
\begin{equation}
    \frac{1}{N\mathcal A}
    \sum_l
    \int_{\mathcal A}d^2r\,
    |{\bf n}_{l+1}-{\bf n}_l|^2
    = \frac{1}{N\mathcal A}
    \sum_l
    \int_{\mathcal A}d^2r\,2(1-{\bf n}_{l+1}\cdot{\bf n}_l)=
    2\left[
    1-\frac{\mathcal C(d)}{\mathcal C(0)}
    \right].
    \label{eq:neighbor_mismatch_C}
\end{equation}
Therefore, the out-of-plane gradient energy density takes the form
\begin{equation}
    f_{g,z}
    =
    \frac{1}{2}
    \Delta\alpha_z P^2,
    \label{eq:fgz_alpha_form}
\end{equation}
where
\begin{equation}
    \Delta\alpha_z
    =
    \frac{2\kappa_z}{d^2}
    \left[
    1-\frac{\mathcal C(d)}{\mathcal C(0)}
    \right].
    \label{eq:alpha_z_C_general}
\end{equation}
Consequently, Eq.~\eqref{eq:alpha_z_C_general} provides a direct relation between
the static interlayer texture correlation and the renormalization of the
polarization stiffness.}

{As discussed in the main text, the polarization structure in the present superlattice exhibits a
periodic modulation along $z$,
correlation function as
\begin{equation}
    \mathcal C(z)
    =
    A
    \left[
    \cos(q_0z+\phi)+C_0
    \right]
    e^{-|z|/\xi},
    \label{eq:dampedosc_general}
\end{equation}
where $q_0=2\pi/\lambda_0$ is the modulation wave vector,
$\lambda_0$ is the superlattice periodicity, $\xi$ is the interlayer
correlation length, $\phi$ is a phase offset, $C_0$ is a nonoscillatory
structural background, and $A$ is an overall amplitude. Substituting Eq.~\eqref{eq:dampedosc_general} into
Eq.~\eqref{eq:alpha_z_C_general}, we obtain
\begin{equation}
    \Delta\alpha_z(\xi)
    =
    \frac{2\kappa_z}{d^2}
    \left[
    1-
    e^{-d/\xi}
    \frac{
    \cos(q_0d+\phi)+C_0
    }{
    \cos\phi+C_0
    }
    \right].
    \label{eq:alpha_z_xi_general}
\end{equation}
The overall amplitude $A$ cancels from the ratio. Consequently, the
gradient-energy correction depends on the spatial organization of the
texture.} 

\begin{figure}[h]
    \centering
    \includegraphics[width=0.5\linewidth]{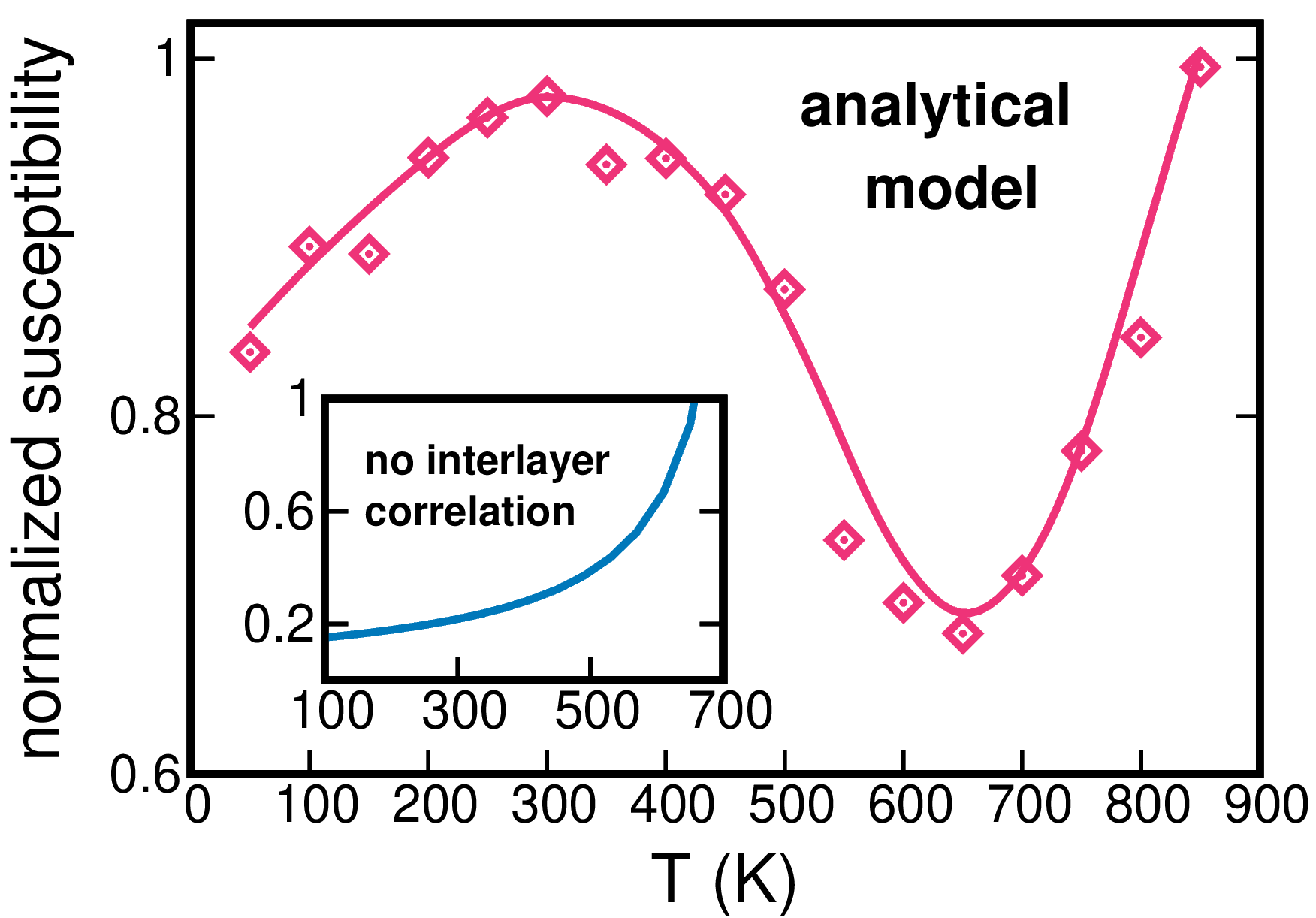}
    \caption{{
    Normalized Dielectric susceptibility obtained from the  effective analytical model  model incorporating skyrmion interlayer correlations based on Eq.~(\ref{eq:chi_xi}). The squares represent the numerical results using the correlation length $\xi(T)$ extracted from the fitted correlation functions and the characteristic temperature ${T}_0\approx760~$K estimated from the ferroelectric background (see Table~\ref{parameterTable}). The solid line is a smooth interpolation of the calculated data points. Inset: dielectric susceptibility in the absence of skyrmions. In this case, the susceptibility exhibits the conventional ferroelectric behavior, increasing monotonically from zero temperature toward the ferroelectric transition temperature $T_c$.
    }}
    \label{fig:SD}
\end{figure}

{Combining the single-layer skyrmion contribution and the interlayer
correlation contribution, the effective free-energy density for the polarization amplitude is
\begin{equation}
    f_{\rm eff}(P)
    =
    \frac{\alpha_{\rm eff}}{2}P^2
    +
    \frac{\beta}{4}P^4
    +
    \frac{\gamma}{6}P^6
    -
    m_zE_zP,
    \label{eq:effective_free_energy_P}
\end{equation} 
where the effective quadratic Landau coefficient incorporating the total texture-induced quadratic
renormalization is 
\begin{equation}
    \alpha_{\rm eff}
    =\alpha+\Delta\alpha_{\rm sk}
    +
    \Delta\alpha_z=
    \alpha
    +
    \kappa_\perp q_{\rm sk}^2
    +
    \frac{2\kappa_z}{d^2}
    \left[
    1-
    e^{-d/\xi}
    \frac{
    \cos(q_0d+\phi)+C_0
    }{
    \cos\phi+C_0
    }
    \right].
    \label{eq:alpha_eff_general}
\end{equation}
The equilibrium condition is
\begin{equation}
    \frac{\partial f_{\rm eff}}{\partial P}
    =
    \alpha_{\rm eff}P
    +
    \beta P^3
    +
    \gamma P^5
    -
    m_zE_z
    =
    0.
    \label{eq:equilibrium_P}
\end{equation}
For a weak probing field, let
\begin{equation}
    P=P_0+\delta P.
\end{equation}
Linearizing Eq.~\eqref{eq:equilibrium_P} gives
\begin{equation}
    \left[
    \alpha_{\rm eff}
    +
    3\beta P_0^2
    +
    5\gamma P_0^4
    \right]
    \delta P
    =
    m_z\delta E_z.
    \label{eq:linearized_P_response}
\end{equation}
Therefore,
\begin{equation}
    \frac{\partial P}{\partial E_z}
    =
    \frac{
    m_z
    }{
    \alpha_{\rm eff}
    +
    3\beta P_0^2
    +
    5\gamma P_0^4
    }.
    \label{eq:dP_dE}
\end{equation}
Since the measured macroscopic polarization is
$\overline P_z=m_zP$, the static dielectric susceptibility is
\begin{align}
    \chi_{zz}
    &=
    \frac{\partial\overline P_z}{\partial E_z}
    =
    m_z\frac{\partial P}{\partial E_z}=
    \frac{
    m_z^2
    }{
    \alpha_{\rm eff}
    +
    3\beta P_0^2
    +
    5\gamma P_0^4
    }.
    \label{eq:chi_general_alphaeff}
\end{align}
Substituting Eq.~\eqref{eq:alpha_eff_general} and Eq.~(\ref{eq:Landau_curvature}), we obtain
\begin{equation}
\chi_{zz}
=
\frac{
m_z^2
}{K(T)+\kappa_\perp q_{\rm sk}^2+\frac{2\kappa_z}{d^2}\big[
1-
e^{-d/\xi}
\frac{
\cos(q_0d+\phi)+C_0
}{
\cos\phi+C_0
}\big]
}.
\label{eq:chi_full_general}
\end{equation}
For large $C_0$ as obtained in the present study and $d/\xi<1$, Eq.~(\ref{eq:chi_full_general}) approximately reduces to
\begin{equation}
\chi_{zz}
=
\frac{m_z^2}
{K(T)+\kappa_\perp q_{\rm sk}^2+{2\kappa_z}/(\xi d)}.
\label{eq:chi_simplified}
\end{equation}
Assuming the conventional Landau form $K(T)=c_{K}({T}_0-T)$ and absorbing the weakly temperature-dependent single-layer skyrmion contribution into a renormalized transition temperature ${T}_0$, one obtains
\begin{equation}
\chi_{zz}(T)\approx
\frac{m_z^2/c_K}
{{T}_0-T+\eta/\xi(T)},
\label{eq:chi_xi}
\end{equation}
where $\eta=2\kappa_z/(c_Kd)$ is an effective coupling constant. Using the correlation length $\xi(T)$ extracted from the fitted correlation functions and estimating ${T}_0$ from the ferroelectric background, the resulting susceptibility is plotted in Fig.~\ref{fig:SD}. As shown, this analytical model model captures the essential features of the temperature dependence obtained from the full phase-field simulations, including the emergence of the broad susceptibility peak.}

{Consequently, the susceptibility is governed by two competing effects: the growth of the correlation length $\xi(T)$, which enhances the dielectric response, and the increasing polarization stiffness $K(T)$, which suppresses fluctuations. The broad maximum in $\chi(T)$ emerges naturally when the enhancement associated with increasing skyrmion correlations is overtaken by the stiffening of the ferroelectric background upon cooling. From the analytical model, the susceptibility peak emerges at approximately $350~\mathrm{K}$, in reasonable agreement with the value of about $400~\mathrm{K}$ obtained from the full phase-field simulations. This peak occurs well below the ferroelectric transition temperature, indicating a hidden crossover associated with the development of skyrmion correlations.}

\subsection{Relaxor-like response}

{For the frequency-dependent dielectric response, a proper description  would require a dynamical framework that treats both the intrinsic polarization dynamics and the collective dynamics of polar skyrmions, including the evolution of interlayer correlations. Here we derive an effective analytical model that explicitly incorporates the contribution of interlayer skyrmion correlations into the frequency-dependent dielectric response. The resulting framework provides an explicit connection between the evolution of the skyrmion texture and the observed ``relaxor-like'' dielectric anomaly.}

{Specifically, we write the polarization field in the $l$-th ferroelectric layer as
\begin{equation}
    {\bf P}_{l}({\bf r},t)
    =
    P(t)\,{\bf n}_{l}({\bf r},t),
    \qquad
    {\bf r}=(x,y),
    \qquad
    |{\bf n}_{l}({\bf r},t)|=1.
    \label{eq:P_texture_dynamic}
\end{equation}
Here $P(t)$ denotes the polarization amplitude, whereas
${\bf n}_{l}({\bf r},t)$ describes the spatially nonuniform skyrmion
texture. The spatially averaged out-of-plane polarization is
\begin{equation}
    \overline{P}_{z}(t)
    =
    \frac{1}{N\mathcal A}
    \sum_{l=1}^{N}
    \int_{\mathcal A} d^2r\,
    P(t)n_{l,z}({\bf r},t)
    =
    m_z(t)P(t),
    \label{eq:Pbar_dynamic}
\end{equation}
where $m_z(t)=\frac{1}{N\mathcal A}\sum_{l=1}^{N}\int_{\mathcal A} d^2r\,n_{l,z}({\bf r},t)$ 
is the geometrical projection factor of the skyrmion texture onto the
direction of the applied field. The electric-field coupling is
therefore
\begin{equation}
    f_E=-m_zE(t)P(t).
    \label{eq:field_coupling_dynamic}
\end{equation}}

{As derived above [Eq.~(\ref{eq:effective_free_energy_P})], the effective total free-energy
density is
\begin{equation}
    f(P,\xi;E)
    =
    \frac{\alpha}{2}P^2
    +
    \frac{\beta}{4}P^4
    +
    \frac{\gamma}{6}P^6
    +
    \frac{\Delta\alpha_{\mathrm{sk}}}{2}P^2
    +
    \frac{\Delta\alpha_z(\xi)}{2}P^2
    -
    m_zEP.
\label{eq:full_free_energy}
\end{equation}
The
dissipative dynamics of $P$ and $\xi$ are described by
\begin{equation}
    \Gamma_P\frac{\partial P}{\partial t}
    =
    -\frac{\partial f}{\partial P},
    \label{eq:LK_P}
\end{equation}
and
\begin{equation}
    \Gamma_\xi\frac{\partial \xi}{\partial t}
    =
    -\frac{\partial f}{\partial \xi},
    \label{eq:LK_xi}
\end{equation}
where $\Gamma_P>0$ and $\Gamma_\xi>0$ are kinetic coefficients.}

{We apply a weak ac field and write
\begin{equation}
    E(t)=E_0+\delta E(t),
    \qquad
    P(t)=P_0+\delta P(t),
    \qquad
    \xi(t)=\xi_0+\delta\xi(t).
    \label{eq:small_fluctuations}
\end{equation}
Using Eq.~\eqref{eq:full_free_energy}, to linear order, the Landau-Khalatnikov equations become
\begin{equation}
    \Gamma_P\delta\dot P
    +
    K_P\delta P
    +
    g\delta\xi
    =
    m_z\delta E,
\label{eq:linear_P}
\end{equation}
\begin{equation}
    \Gamma_\xi\delta\dot\xi
    +
    K_\xi\delta\xi
    +
    g\delta P
    =
    0.
\label{eq:linear_xi}
\end{equation}
Here, the polarization stiffness is
\begin{equation}
    K_P=\left.
    \frac{\partial^2f}{\partial P^2}
    \right|_{P_0,\xi_0}
    =
    \alpha
    +
    3\beta P_0^2
    +
    5\gamma P_0^4
    +
    \Delta\alpha_{\mathrm{sk}}
    +
    \Delta\alpha_z(\xi_0).
\label{eq:KP}
\end{equation}
The mixed coupling is
\begin{equation}
    g=
    \left.
    \frac{\partial^2f}{\partial P\partial\xi}
    \right|_{P_0,\xi_0}
    =
    P_0\Delta\alpha_z'(\xi_0).
\label{eq:g_general}
\end{equation}
The interlayer-correlation stiffness is
\begin{equation}
    K_\xi=
    \left.
    \frac{\partial^2f}{\partial\xi^2}
    \right|_{P_0,\xi_0}=
    \frac{P_0^2}{2}
    \Delta\alpha_z''(\xi_0)
    =
    \frac{\kappa_zRP_0^2}{d}
    e^{-d/\xi_0}
    \frac{2\xi_0-d}{\xi_0^4}.
\label{eq:Kxi_explicit}
\end{equation}}

{We use the Fourier convention
\begin{equation}
    \delta P(t)=\delta P(\omega)e^{-i\omega t},
    \qquad
    \delta\xi(t)=\delta\xi(\omega)e^{-i\omega t},
    \qquad
    \delta E(t)=\delta E(\omega)e^{-i\omega t}.
    \label{eq:Fourier_convention}
\end{equation}
Then Eqs.~\eqref{eq:linear_P} and \eqref{eq:linear_xi} become
\begin{equation}
    \left(
    K_P-i\omega\Gamma_P
    \right)\delta P
    +
    g\delta\xi
    =
    m_z\delta E,
    \label{eq:frequency_P}
\end{equation}
and
\begin{equation}
    g\delta P
    +
    \left(
    K_\xi-i\omega\Gamma_\xi
    \right)\delta\xi
    =
    0.
    \label{eq:frequency_xi}
\end{equation}
Solving the second equation gives
\begin{equation}
    \delta\xi(\omega)
    =
    -
    \frac{g}{K_\xi}
    \frac{1}{1-i\omega\tau_\xi}
    \delta P(\omega).
\label{eq:xi_response_Debye}
\end{equation}
where $\tau_\xi
    ={\Gamma_\xi}/{K_\xi}$. At low frequency, $\omega\tau_\xi\ll1$, the interlayer correlation follows
the polarization quasi-statically. At high frequency,
$\omega\tau_\xi\gg1$, the interlayer correlation is dynamically frozen.}

{Substituting Eq.~\eqref{eq:xi_response_Debye} into
Eq.~\eqref{eq:frequency_P}, we obtain
\begin{equation}
    \left[
    K_P-i\omega\Gamma_P
    -
    \frac{g^2/K_\xi}
    {1-i\omega\tau_\xi}
    \right]\delta P
    =
    m_z\delta E.
    \label{eq:effective_P_equation}
\end{equation}
Using Eq.~\eqref{eq:effective_P_equation}, $\delta\overline P_z(\omega)=m_z\delta P(\omega)$, the dielectric susceptibility can first be written as
\begin{equation}
\chi_{zz}(\omega)
=
\frac{
m_z^2\left(1-i\omega\tau_\xi\right)
}{
\left(K_P-i\omega\Gamma_P\right)
\left(1-i\omega\tau_\xi\right)
-
g^2/K_\xi
}.
\label{eq:chi_full_relaxation}
\end{equation}
We assume
that the local polarization amplitude relaxes much faster than the
skyrmion interlayer correlation. In the frequency range of interest, this
condition is expressed as
\begin{equation}
\omega\Gamma_P\ll K_P,
\label{eq:fast_P_condition}
\end{equation}
and hence, the intrinsic polarization damping term can then be neglected, yielding 
\begin{equation}
\chi_{zz}(\omega)
\simeq
\frac{
m_z^2\left(1-i\omega\tau_\xi\right)
}{
K_P\left(1-i\omega\tau_\xi\right)
-
g^2/K_\xi
}.
\label{eq:chi_fast_P}
\end{equation}}

{We define the relaxed polarization stiffness as
\begin{equation}
K_{\mathrm r}
=
K_P-\frac{g^2}{K_\xi},
\label{eq:Kr_definition}
\end{equation}
and the effective relaxation time is defined as
\begin{equation}
\tau_{\mathrm{eff}}
=
\frac{K_P}{K_{\mathrm r}}\tau_\xi
=
\frac{K_P}
{K_P-g^2/K_\xi}
\tau_\xi.
\label{eq:tau_effective}
\end{equation}
Then, Eq.~\eqref{eq:chi_fast_P} can be rearranged into the conventional Debye form
\begin{equation}
\chi_{zz}(\omega)
=
\chi_\infty
+
\frac{
\chi_0-\chi_\infty
}{
1-i\omega\tau_{\mathrm{eff}}
}.
\label{eq:chi_Debye}
\end{equation}
Therefore, the dynamical reorganization of the skyrmion interlayer correlation produces the same mathematical form as the dielectric response
of a relaxational ferroelectric.}

{Neglecting the weak temperature dependence of $\chi_\infty$, the real
part of the susceptibility is
\begin{equation}
\chi_{zz}'(\omega,T)
=
\chi_\infty
+
\frac{
\Delta\chi(T)
}{
1+\omega^2\tau_{\mathrm{eff}}^2(T)
},
\qquad
\Delta\chi(T)
=
\chi_0(T)-\chi_\infty.
\label{eq:chi_real_existing_peak}
\end{equation}
Let $T_0$ denote the position of the static maximum, such that
\begin{equation}
\Delta\chi'(T_0)=0,
\qquad
\Delta\chi''(T_0)<0.
\label{eq:static_peak_condition}
\end{equation}}

{At finite frequency, the peak temperature $T_{\mathrm p}(\omega)$ is
determined by
\begin{equation}
\left.
\frac{\partial\chi_{zz}'(\omega,T)}
{\partial T}
\right|_{T=T_{\mathrm p}}
=
0.
\label{eq:finite_frequency_peak_condition}
\end{equation}
Differentiating Eq.~\eqref{eq:chi_real_existing_peak} gives
\begin{equation}
\frac{
\Delta\chi'(T_{\mathrm p})
}{
\Delta\chi(T_{\mathrm p})
}
=
\frac{
2\omega^2
\tau_{\mathrm{eff}}(T_{\mathrm p})
\tau_{\mathrm{eff}}'(T_{\mathrm p})
}{
1+
\omega^2\tau_{\mathrm{eff}}^2(T_{\mathrm p})
}.
\label{eq:peak_shift_general}
\end{equation}}

\begin{figure}[h]
    \centering
    \includegraphics[width=0.5\linewidth]{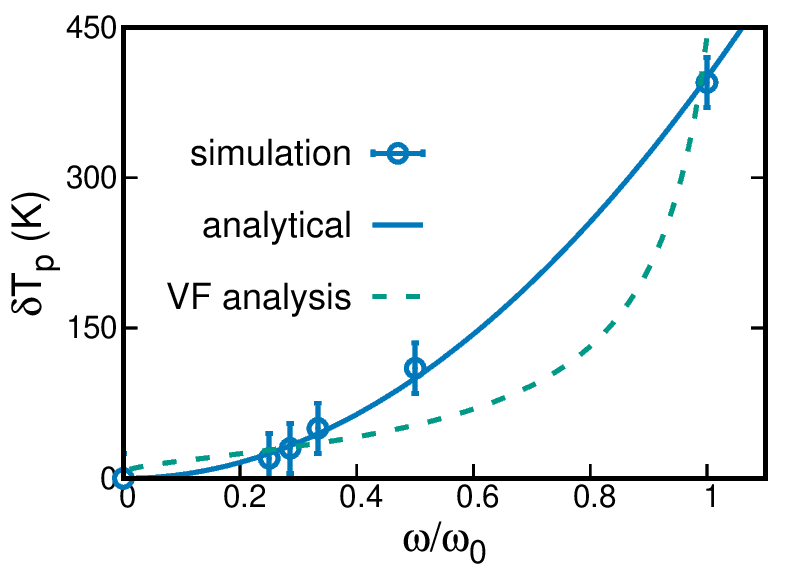}
\caption{Frequency dependence of the susceptibility peak-temperature shift,
    $\delta T_p(\omega)=T_p(\omega)-T_p(0)$, extracted from the full phase-field
    simulations. The symbols represent the simulation results, the solid
    curve shows the quadratic fit $\delta T_p(\omega)\propto\omega^2$, and 
   the dashed curve shows the best fit to the VF relation. 
    Here, $\omega_0=1/t_0$.}
    \label{fig:SE}
\end{figure}

{To obtain the peak displacement explicitly, we write
\begin{equation}
T_{\mathrm p}(\omega)=T_0+\delta T_{\mathrm p}.
\end{equation}
Expanding near $T_0$, $\Delta\chi'(T_{\mathrm p})
\simeq
\Delta\chi''(T_0)\delta T_{\mathrm p}$, and then, using Eq.~(\ref{eq:peak_shift_general}), 
the leading peak shift is then
\begin{equation}
\delta T_{\mathrm p}
=
\frac{
\Delta\chi(T_0)
}{
\Delta\chi''(T_0)
}
\frac{
2\omega^2
\tau_{\mathrm{eff}}(T_{\mathrm p})
\tau_{\mathrm{eff}}'(T_{\mathrm p})
}{
1+
\omega^2\tau_{\mathrm{eff}}^2(T_{\mathrm p})
}.
\label{eq:peak_shift_explicit}
\end{equation}
For a texture dynamics that slows down upon cooling, one has the relaxation time $\tau'_{\rm eff}(T_P)<0$. Together with $\Delta\chi(T)>0$ and $\Delta\chi''(T_0)<0$, this gives rise to 
\begin{equation}
\delta T_{\mathrm p}>0.
\end{equation}
Consequently, the maximum of the finite-frequency susceptibility shifts toward higher temperatures as the probing frequency increases, consistent with the full phase-field simulations. Figure~\ref{fig:SE} presents the frequency dependence of the peak temperature from the full phase-field simulation. As shown, the peak shift, $\delta T_p(\omega)$, exhibits a clear quadratic dependence on frequency, $\delta T_p(\omega)\propto \omega^2$, in excellent agreement with the analytical model.}

It should be emphasized that while the polar structures shown in Fig. 2(a) exhibit exhibit variations in size and shape, they are fundamentally different from the polar nanoregions (PNRs) found in conventional relaxor ferroelectrics. In traditional relaxors, PNRs originate from quenched chemical disorder and the associated {\sl random} electric fields. Their existence, dynamics and spatial feature are therefore intimately tied to compositional inhomogeneity. The polar structures in our simulations emerge in a chemically ordered superlattice. Their formation is driven by the competition among ferroelectric, elastic, electrostatic, and gradient energies inherent to the superlattice architecture, and the obtained nanoscale polar objects, polar skyrmions, possess a {\sl fixed} nontrivial topological structure.

{
We also note that the absence of quenched chemical disorder or explicitly
introduced glassy degrees of freedom does not, by itself, preclude the
emergence of Vogel-Fulcher (VF)-type dynamics. Glassy dynamics can emerge
collectively from a sufficiently complex free-energy landscape even when no
explicit glassy term is introduced in the underlying free-energy functional.
In particular, multiple metastable configurations can provide a glassy-like
configurational landscape whose thermally activated fluctuations naturally
lead to collective slowing down and VF-type dynamics. The present system, however, does not exhibit such a mechanism. Unlike
ordinary dipolar textures that can continuously reorganize and transform
through fluctuations, the skyrmions are constrained to preserve their
topological character,
thereby shifting the dominant collective dynamics from configurational
fluctuations among competing textures to the formation and dissolution of
correlations among the topological textures, as also reflected in the
theoretical derivation above. To substantiate this distinction, we explicitly examine whether the frequency
dependence obtained from our phase-field simulations can be described by the
conventional VF relation,
relation is
\begin{equation}
\omega
=
\omega_c
\exp\left[
-\frac{E_a}{k_B(T_P-T_f)}
\right],
\end{equation}
which gives
\begin{equation}
T_P(\omega)-T_f
=
\frac{E_a/k_B}
{\ln(\omega_c/\omega)}.
\end{equation}
In contrast, the dynamical response associated with the interlayer skyrmion
correlations derived above gives
\begin{equation}
\delta T_P(\omega)=
T_P(\omega)-T_P(0)
\propto \omega^2,
\qquad \text{in the low-frequency regime}.
\end{equation}
As shown in Fig.~\ref{fig:SE}, the peak temperatures extracted directly from the full
phase-field simulations follow this quadratic frequency dependence predicted
by the dynamical framework of interlayer skyrmion correlations, whereas the
VF form does not reproduce the observed frequency dependence. Therefore, the dielectric response observed in the present system should be regarded as relaxor-like in its phenomenology, while its frequency-dependent peak shift is governed by correlation dynamics rather than by VF-type glassy freezing.
}

\bibliography{reference.bib}